\documentclass[a4paper,twocolumn,11pt,accepted=2025-03-20]{quantumarticle}
\pdfoutput=1

\usepackage[square,numbers]{natbib}

\usepackage[table,dvipsnames]{xcolor}
\usepackage[utf8]{inputenc}
\usepackage[english]{babel}
\usepackage[T1]{fontenc}
\usepackage[ruled,vlined]{algorithm2e}
\usepackage{flowchart}
\usepackage{fontawesome5}
\usepackage{svg}
\usepackage{authblk}
\usepackage{subcaption}
\usepackage{bm}
\usepackage{pifont}
\newcommand{\cmark}{{\color{green}\ding{51}}}%
\newcommand{\xmark}{{\color{red}\ding{55}}}%

\usepackage{cancel}
\usepackage{soul}
\usepackage[normalem]{ulem}
\newcommand{\stkout}[1]{\ifmmode\text{\sout{\ensuremath{#1}}}\else\sout{#1}\fi}
\newcommand\hcancel[2][black]{\setbox0=\hbox{$#2$}%
\rlap{\raisebox{.45\ht0}{\textcolor{#1}{\rule{\wd0}{1pt}}}}#2}

\usepackage{tikz}
\usetikzlibrary{backgrounds,fit,decorations.pathreplacing}

\tikzstyle{process} = [rectangle, draw, fill=blue!20, text centered, rounded corners, minimum height=2em]

\usetikzlibrary{calc}

\usepackage{amsmath}
\usepackage{amssymb}
\usepackage{bbm}
\usepackage{physics}

\usepackage{hyperref}

\usepackage{csquotes}

\usepackage{centernot}
\usepackage{mathtools}
\usepackage{stmaryrd}

\makeatletter
\newcommand{\xMapsto}[2][]{\ext@arrow 0599{\Mapstofill@}{#1}{#2}}
\def\Mapstofill@{\arrowfill@{\Mapstochar\Relbar}\Relbar\Rightarrow}
\makeatother

\newcommand{\C}[0]{\mathbb{C}}

\usepackage{listings}

\definecolor{codegreen}{rgb}{0,0.6,0}
\definecolor{codegray}{rgb}{0.5,0.5,0.5}
\definecolor{codepurple}{rgb}{0.58,0,0.82}
\definecolor{backcolour}{rgb}{0.95,0.95,0.92}

\lstdefinestyle{mystyle}{
    backgroundcolor=\color{backcolour},
    commentstyle=\color{codegreen},
    keywordstyle=\color{magenta},
    numberstyle=\tiny\color{codegray},
    stringstyle=\color{codepurple},
    basicstyle=\ttfamily\footnotesize,
    breakatwhitespace=false,
    captionpos=b,
    keepspaces=true,
    numbers=left,
    numbersep=5pt,
    showspaces=false,                
    showstringspaces=false,
    showtabs=false,
    tabsize=2
}

\usepackage{float}
\usepackage{listings}

\newfloat{lstfloat}{htbp}{lop}
\floatname{lstfloat}{Listing}

\ifdefined\highlighted
    
\else
    
    \renewcommand{\sout}[1]{}
    \renewcommand{\stkout}[1]{}
    \renewcommand{\st}[1]{}
    \renewcommand{\hcancel}[1]{}
\fi

\newcommand{\ann}{\hat{a}}
\newcommand{\hx}{\hat{x}}
\newcommand{\hp}{\hat{p}}

\title{Piquasso: A Photonic Quantum Computer Simulation Software Platform}

\author[1, 2]{Zoltán Kolarovszki}
\email{kolarovszki.zoltan@wigner.hun-ren.hu}
\author[3]{Tomasz Rybotycki}
\author[4]{Péter Rakyta}
\author[1, 2]{Ágoston Kaposi}
\author[2, 5]{Boldizsár Poór}
\author[1, 2, 6]{Szabolcs Jóczik}
\author[1,4]{Dániel T. R. Nagy}
\author[2]{Henrik Varga}
\author[1, 7]{Kareem H. El-Safty}
\author[2]{Gregory Morse}
\author[3,8]{Micha{\l}  Oszmaniec}
\author[2]{Tamás Kozsik}
\author[1, 2, 10]{Zoltán Zimborás}
\email{zimboras.zoltan@wigner.hun-ren.hu}

\affil[1]{Quantum Computing and Quantum Information Research Group, HUN-REN Wigner Research Centre for Physics, Konkoly–Thege Miklós út 29-33, H-1525 Budapest, Hungary}
\affil[2]{Department of Programming Languages and Compilers, E\"otv\"os  Lor\'and  University, Pázmány Péter sétány 1/a, H-1117 Budapest, Hungary}
\affil[3]{Center for Theoretical Physics, Polish Academy of Sciences, Al. Lotników 32/46, 02-668 Warszawa, Poland}
\affil[4]{Department of Physics of Complex Systems, E\"otv\"os  Lor\'and  University, Pázmány Péter sétány 1/a, H-1117 Budapest, Hungary}
\affil[5]{Quantinuum, 17 Beaumont Street, Oxford, OX1 2NA, United Kingdom}
\affil[6]{Robert Bosch Kft., Gyömrői út 104., H-1103 Budapest, Hungary}
\affil[7]{Department of Computer Engineering, Technical University of Munich, Arcisstraße 21, 80333 München, Germany}
\affil[9]{NASK National Research Institute, Kolska 12,  01-045 Warsaw,
            Poland}
\affil[10]{Algorithmiq Ltd, Kanavakatu 3C 00160 Helsinki, Finland}

\makeatletter
\newcommand{\verbatimfont}[1]{\renewcommand{\verbatim@font}{\ttfamily#1}}
\makeatother
\makeatletter
\preto{\@verbatim}{\topsep=0pt \partopsep=0pt }
\makeatother

\newcommand{\specialcell}[2][c]{%
\begin{tabular}[#1]{@{}c@{}}#2\end{tabular}}

\date{29th February, 2024}

\begin{document}

\maketitle

\begin{abstract}
    We introduce the Piquasso quantum programming framework, a full-stack open-source software platform for the simulation and programming of photonic quantum computers.
    Piquasso can be programmed via a high-level Python programming interface enabling users to perform efficient quantum computing with discrete and continuous variables.
    Via optional high-performance C++ backends, Piquasso provides state-of-the-art performance in the simulation of photonic quantum computers.
    The Piquasso framework is supported by an intuitive web-based graphical user interface where the users can design quantum circuits, run computations, and visualize the results.
\end{abstract}

\section{Introduction} \label{sec:introduction}
    In the last decade, there has been great progress in creating quantum computer prototypes. 
    Among the many proposals, the relevance of photonic quantum computers increased due to recent demonstrations of possible photonic quantum advantage schemes~\cite{photonicadvantage1, zhong2021phase,madsen2022quantum} and the development of feasible fault-tolerant quantum computation methods~\cite{bartolucci2023fusion,bombin2023logical,bourassa2021blueprint}.

    In parallel with the progress on quantum hardware prototypes, the need for quantum computer simulators, and generally quantum software, has been steadily increasing. There are manifold reasons why classical simulators are needed: Current quantum devices are still noisy, so it is instructive to compare experimental results with the ideal noiseless outcomes obtained from the simulator. Moreover, one can study with noiseless simulators the performance of new heuristic algorithms, e.g., quantum neural networks or variational quantum eigensolvers. In addition to this, by implementing flexible noise models in the simulator, one could test the noise tolerance of quantum algorithms and evaluate the usefulness of different error mitigation or even error correction schemes.

    Consequently, in recent years plenty of quantum computing simulation platforms have been developed~\cite{fingerhuth2018open}. However, most of these focus on qubit-based quantum computing.
    Much less development has been done for photonic quantum computation. 
    Strawberry Fields~\cite{sf:2019}, developed by Xanadu, is an open-source quantum programming platform built using Python~\cite{python}, which contains a simulator and can also serve as an interface for existing hardware, e.g., the Borealis chip~\cite{madsen2022quantum}. We should also mention here Perceval~\cite{heurtel2023perceval}, Bosonic Qiskit~\cite{stavenger2022bosonic}, GraphiQ~\cite{lin2024graphiq}, MrMustard~\cite{mrmustard} and BosonSampling.jl~\cite{bosonsampling_jl}.
    Perceval developed by Quandela is also a framework for simulating optical elements, however, it does not aim to treat continuous-variable models of photonic quantum computation.
    Bosonic Qiskit is also capable of simulating optical elements, however, it is primarily aimed at modeling hybrid quantum computation containing both bosonic and qubit-based objects; and is less emphasized for photonic systems, it naturally lacks some gates specific for photonic quantum computation such as the Kerr and Cross-Kerr gates. GraphiQ is a library dedicated to the simulation of photonic graph states, also selecting a different application domain.
    Finally, MrMustard is dedicated for differentiable simulation of Gaussian circuits using the Bargmann representation of Gaussian states, which is not yet implemented in Piquasso.
    In Table~\ref{table:feature_table}, we summerize the features of the relevant software packages admitting similar applications as Piquasso is designed for.

    Considering the narrow selection of photonic quantum computer simulators, we developed a new photonic quantum computer simulator software called Piquasso (PhotonIc QUAntum computer Simulator SOftware). Creating a new framework is beneficial for several reasons, for example: (i) Rethinking existing classical simulations might help in the development of more efficient classical algorithms (see our torontonian implementation~\cite{kaposi2022polynomial} that was also taken over by Strawberry Fields).
    (ii) New design choices can be implemented that enable practical features different from the existing ones. In our case, these are, e.g., the repeatability of the simulations via seeding, the increase in the simulable number of modes by the choice of Fock space truncation, and the possibility of replacing or extending default calculations via plugins (e.g., see Sec.~\ref{sec:extending}).
    (iii) It can be useful to have multiple simulators testing photonic hardware.
    (iv) One can enable different numerical computing frameworks, e.g., TensorFlow~\cite{tensorflow2015-whitepaper} or JAX~\cite{jax2018github}.
    Piquasso has found many applications recently, such as simulation of non-deterministic gates~\cite{czaban2024suppressingphotondetectionerrors}, continuous-variable Born-machines~\cite{cvbm_article} and quantum machine learning~\cite{nagy2024hybridquantumclassicalreinforcementlearning}.

    The article is structured as follows: Sec.~\ref{sec:overview} shortly summarizes the goals and main features of Piquasso, while 
    Sec.~\ref{sec:platform} describes the hierarchy of the Piquasso platform and gives several code examples.
    Sec.~\ref{sec:benchmarks} briefly discusses the computational performance of Piquasso.
    The web user interface of Piquasso is presented in Sec.~\ref{sec:ui}.
    Finally, Appendix~\ref{app:basics} summarizes the primary concepts and basic calculations in photonic quantum computing.

\section{Goals} \label{sec:overview}
    Our main aim with Piquasso is to enhance research on quantum optical computation by providing an accessible platform for the simulation of photonic quantum computers. On the one hand, we intend to provide a user-friendly and easy-to-extend system. On the other hand, we also focus on boosting the efficiency of the computations executed during the simulation.
    With the development of Piquasso, we seek to fulfill the following criteria:
    \begin{itemize}
        \item \textbf{Speed}: In the PiquassoBoost extension, calculations of several classically computationally hard quantities are rewritten in C/C++ in order to enable more efficient calculations.
        \item \textbf{Extendability}: Piquasso comes with a public interface to enable users to customize instructions and calculations or even to create new simulators and quantum state datatypes.
        \item \textbf{Repeatability}: For non-deterministic computations, one can extract the seed of the random number generation, save it, and repeat the same simulation later.
        \item \textbf{Intuitive user interface}: In the user interface, one can specify high-level instructions like an interferometer, a general Gaussian gate, or even a graph embedding.
        \item \textbf{Quantum Machine Learning support}: TensorFlow is an end-to-end machine learning platform~\cite{tensorflow2015-whitepaper}, which supports automatic differentiation. The simulation of pure Fock states can be performed using TensorFlow as a calculation backend. This way, one can compute the gradient of certain simulations and use it for machine learning purposes. Moreover, Piquasso also supports the JAX machine learning framework for performing backend calculations.
    \end{itemize}

    \begin{table*}[ht]
    \begin{center}
    \hspace{0.5cm}
    \bgroup
    \def\arraystretch{1.0}
    \begin{tabular}{|c|c|c|c|c|c|c|c|}
            \hline
            & \specialcell{Gaussian-state\\based simulations} & \specialcell{Fock-space\\based simulations} &  \specialcell{Efficient\\BS} & \specialcell{Efficient\\GBS} & \specialcell{PQC with\\differentiabilty} \\ \hline
        Strawberry Fields & \cmark &  \cmark &  \xmark & \cmark & \cmark \\ \hline
        MrMustard & \cmark & \xmark &  \xmark & \xmark & \cmark \\ \hline
        Perceval  & \xmark  & \cmark & \cmark & \xmark & \xmark \\ \hline
        Piquasso  & \cmark  & \cmark & \cmark & \cmark & \cmark \\ \hline
    \end{tabular}
    \egroup
    \end{center}
    \caption{
        Comparison of different photonic quantum computer simulation frameworks. Notation: BS -- Boson Sampling, GBS -- Gaussian Boson Sampling, PQC -- Parametric Quantum Circuits.
    }\label{table:feature_table}
    \end{table*}

\section{Piquasso Platform} \label{sec:platform}
    
    Piquasso is an open source software, designed to provide a simple, accessible interface, which enables users to extend the built-in simulators or define new ones. It is written using Python~\cite{python}, a language already familiar with scientific computations.
    The source code is made available at~\cite{piquasso}, and the documentation is published at~\cite{documentation}.

    Using \lstinline{pip}, Piquasso can simply be installed from \href{https://pypi.org/}{PyPI} using
    \vspace{1em}
    \verbatimfont{\small}%
    \begin{verbatim}
pip install piquasso
    \end{verbatim}
    As a dependency, the NumPy scientific computing package is installed~\cite{numpyarray}, which is also a helpful companion for writing programs for scientific computing purposes.
    It is important to note, that using Piquasso only assumes basic knowledge of Python, as illustrated by the Code snippet~\ref{code:basicexample}.
    
    \subsection{Hierarchy}
\begin{figure*}[t]
        
            \centering
    
            \begin{tikzpicture}[>=stealth, thick]
            
            \node (Co) at (0,0) [draw, process, align=flush center] {\lstinline{Config}};
            
            \node (St) at (2,0) [draw, process, align=flush center]  {\lstinline{State}};
            
            \node (Si) at (1,-2) [draw, process, align=flush center] 
            {\lstinline{Simulator}};
            
            \node (Pre) at (5, 0) [draw, process, align=flush center] 
            {\lstinline{Preparation}};
    
            \node (Ga) at (7, 0) [draw, process, align=flush center] 
            {\lstinline{Gate}};
            
            \node (Me) at (9, 0) [draw, process, align=flush center] 
            {\lstinline{Measurement}};
            
            \node (Ch) at (11.2, 0) [draw, process, align=flush center] 
            {\lstinline{Channel}};
            
            \node (Pr) at (8, -2) [draw, process, align=flush center] 
            {\lstinline{Program}};

            \node (Ex) at (4.5, -4) [draw, process, align=flush center] 
            {\lstinline{Simulator.execute}};
            
            \node (Re) at (4.5, -6) [draw, process, align=flush center] 
            {\lstinline{Result}};
            
            \draw[->] (Co) -- (Si);
            \draw[->] (St) -- (Si);
            
            \draw[dashed, ->] (Co) -- (St);
            
            \draw[->] (Pre) -- (Pr);
            \draw[->] (Ga) -- (Pr);
            \draw[->] (Me) -- (Pr);
            \draw[->] (Ch) -- (Pr);
            
            \draw[->] (Si) -- (Ex);
            \draw[->] (Pr) -- (Ex);
            
            \draw[->] (Ex) -- (Re);
    
            \end{tikzpicture}
            
            \caption{Schematic diagram of executing a Piquasso program. The \lstinline{Config} and (optional) initial \lstinline{State} object should be specified to the \lstinline{Simulator}. The dotted line between \lstinline{Config} and \lstinline{State} indicates that \lstinline{State} also depends on the \lstinline{Config} for calculations after execution, but it is generally injected into \lstinline{State} by the \lstinline{Simulator}. The \lstinline{Instruction} instances should be specified for the single \lstinline{Program} instance. Then the \lstinline{Program} instance should be specified to \lstinline{Simulator.execute} to perform the calculations and yield a \lstinline{Result} instance.}
    
            \label{fig:schematic}
        \end{figure*}
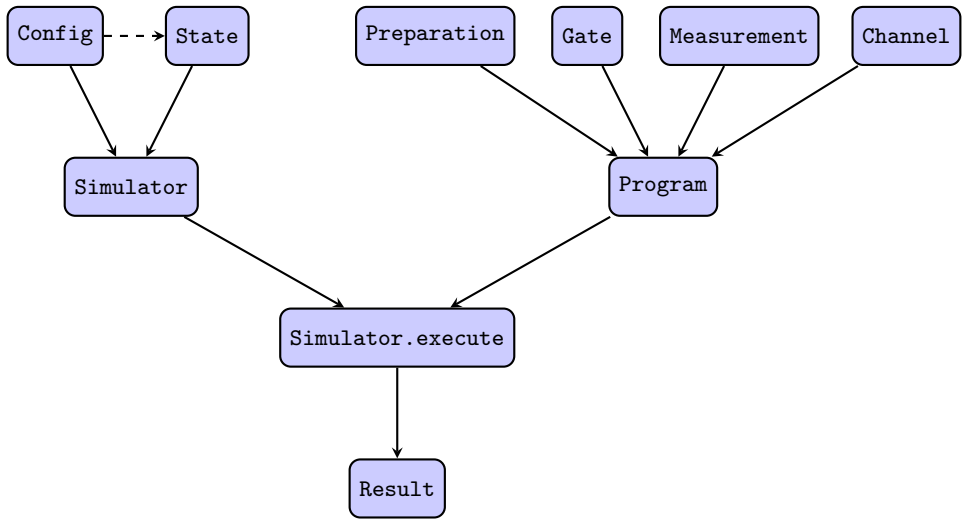

        The schematic dependence between Piquasso objects is illustrated in Fig.~\ref{fig:schematic}. Piquasso is structured in a way that a single \lstinline{Program} instance only contains \lstinline{Preparation}, \lstinline{Gate},  \lstinline{Measurement} and \lstinline{Channel} instances. Strictly speaking, no other information is needed in the program definition. To specify the instruction, the following syntaxes could be followed:
        \begin{lstlisting}[language=Python]
# Single instruction on RHS
pq.Q([MODES]) | [INSTR]([PARAMS])

# Single instruction on LHS
[INSTR]([PARAMS]) | pq.Q([MODES])

# Multiple instructions on RHS
pq.Q([MODES]) | (
    [INSTR_1]([PARAMS])
    | ...
    | [INSTR_N]([PARAMS])
)
\end{lstlisting}    
        where 
        \begin{itemize}
            \item \lstinline{[MODES]} is a sequence of non-negative integers representing the modes on which the instruction should act. As a syntactic sugar, it is permitted to write \lstinline{all} to specify all modes if applicable, or to leave it out entirely;
            \item \lstinline{[INSTR]},  \lstinline{[INSTR_1], ..., [INSTR_M]} are instruction classes, by which preparations, gates, channels, or measurements can be defined;
            \item \lstinline{[PARAMS]} are the parameters specific for each instruction.
        \end{itemize}
    
        The \lstinline{Simulator} instance contains the \lstinline{Config} instance and the \lstinline{State} instance. \lstinline{Config} contains the necessary data to perform simulation (e.g., Planck constant, Fock space cutoff), and the \lstinline{State} instance holds the representation of the quantum state described in Sec.~\ref{sec:states}.
        
        After the simulation is executed, the \lstinline{result} object contains all the samples under \lstinline{result.samples} and the resulting quantum state under \lstinline{result.state}.

\begin{lstlisting}[
            language=Python,
            float=*,
            label={code:basicexample},
            caption={Basic Piquasso code example. On line 5, the program definition starts with a \lstinline{with} statement that will result in a \lstinline{Program} instance. In this block, all the instructions should be specified. The \lstinline{pq.Q} class is used to specify the modes on which the instructions on the other side of the \lstinline{|} or \lstinline{__or__} operator are supposed to act. After the program definition, a \lstinline{GaussianSimulator} instance is created with two modes, and the result is acquired by executing \lstinline{program} with \lstinline{simulator}. The execution parameter \lstinline{shots=1000} is specified in order to acquire 1000 samples from the measurement.
            }
        ]
import numpy as np
import piquasso as pq

# Beginning of program definition
with pq.Program() as program:
    # Initializing vacuum state
    pq.Q(all)  | pq.Vacuum() 

    # Quantum Gates
    pq.Q(0)    | pq.Displacement(r=1.0)
    pq.Q(1)    | pq.Displacement(r=1.0)
    pq.Q(0)    | pq.Squeezing(r=0.1, phi=np.pi / 3)
    pq.Q(0, 1) | pq.Beamsplitter(theta=np.pi / 3, phi = np.pi / 4)
    
    # Measurement
    pq.Q(0, 1) | pq.ParticleNumberMeasurement()

# Choosing a simulator
simulator = pq.GaussianSimulator(d=2)

# Execution
result = simulator.execute(program, shots=1000)
\end{lstlisting}
    
    \subsection{Built-in simulators} \label{ssec:simulators}
        Piquasso allows the simulation of special scenarios, such as computation using solely Gaussian states or pure Fock states,
        since this yields benefits in execution time and memory usage.
        For example, considering the case when only pure Gaussian states and gates are used in the photonic circuit, one might consider using the \lstinline{GaussianSimulator}, which is more efficient than the \lstinline{PureFockSimulator} for several use cases.

        \subsubsection{Fock simulators}  \label{sec:focksimulator}
           Piquasso supports a separate simulator dedicated to working with quantum states in the Fock representation. In this simulator, the states are represented in the occupation number basis (see Appendix~\ref{app:basics}). When no mixed states are required during the simulation, it is sufficient to use the \lstinline{PureFockSimulator} class instead of the more general \lstinline{FockSimulator}. An example usage of the \lstinline{PureFockSimulator} class is shown in Code snippet~\ref{code:purefockexample}.

           Naturally, Fock simulators have a high memory usage due to storing the state vector or density matrix. The Fock space (see, Eq.~\eqref{eq:fockspace} in Appendix~\ref{app:basics})  has to be truncated with a cutoff particle number $c$ determined by the user, i.e., the simulator calculates in the space defined by
            \begin{equation} \label{eq:truncated_fock_space}
                \mathcal{F}^c_{B}(\C^d) =  \bigoplus_{\lambda=0}^{c-1}  \, \left( \C^d \right)^{\vee \lambda}.
            \end{equation}
            The cutoff $c$ sets a limit on the total particle number in the system, which is more memory efficient than the limit on the particle number on each mode, as will be discussed in detail in Sec.~\ref{sec:focksimulatorlosses}.

        \subsubsection{Pure Fock simulator with TensorFlow support}
            A substantial part of Quantum Machine Learning algorithms are developed for photonic architectures. Consequently, we gave special attention to enabling the differentiation of photonic circuits.

            Piquasso uses the automatic differentiation capabilities of TensorFlow~\cite{tensorflow2015-whitepaper}. 
            When simulating a photonic circuit, the Jacobian of the resulting state vector after applying the layers consisting of optical gates can automatically be differentiated.
            However, in TensorFlow Eager mode, custom gradients can be implemented to increase performance instead of using the automatic differentiation of built-in TensorFlow functions.
            Notably, the Jacobians of gate matrices corresponding to linear optical gates can be calculated~\cite{miatto2020fast}.
            Moreover, one can also differentiate the operation of applying a gate to a state vector, which is implemented in Piquasso.
    
            In Piquasso, automatic differentiation is implemented through \lstinline{TensorflowConnector}, where the simulation is restricted to pure Fock states. An example usage can be seen in Code snippet~\ref{code:tf_general} implementing a CVQNN layer for a single mode. For multiple modes, the circuit definition is similar but less concise.

        \subsubsection{Gaussian simulator}
            The Gaussian simulator represents Gaussian quantum states in the phase space formalism, i.e., the states are stored via their mean vectors and their covariance matrices according to Eq.~\eqref{eq:meancov} in Appendix~\ref{app:basics}.
            An example usage of the \lstinline{GaussianSimulator} class is shown in Code snippet~\ref{code:gaussianexample}.
            
            The state inside the Gaussian simulator is evolved by
            \begin{align}
                \mu &\mapsto S \mu, \nonumber \\
                \sigma &\mapsto S \sigma S^T,
            \end{align}
            where $S$ is the symplectic representation corresponding to the unitary evolution and relates to $S_{(c)}$ from Eq.~\eqref{eq:symplectic_unitary} in Appendix~\ref{app:basics}.
    
            The supported gates are the ones with only up to quadratic terms in their Hamiltonians.
            Note that while non-Gaussian measurements execute the sampling during Gaussian simulations, the final state after the measurement is not calculated since it is not Gaussian in general.
    
            The Gaussian simulator is generally faster and more precise than performing the same simulation with the Fock simulator from Sec.~\ref{sec:focksimulator}. It also has a lower memory usage, since the mean vector and covariance matrix in Eq.~\eqref{eq:meancov} from Appendix~\ref{app:basics} requires considerably less memory than a density matrix over the Fock space.

            When executing Gaussian measurements, the Gaussian simulator uses high-complexity matrix functions to calculate the probabilities corresponding to different measurement outcomes, as described in Appendix~\ref{app:particle_number} and Appendix~\ref{app:threshold}. The efficient calculation of these functions is crucial for many applications, hence, Piquasso implements cutting-edge algorithms.
            The speedup of Piquasso (version 5.0.0) over the TheWalrus (version 0.21.0) is presented in Sec.~\ref{sec:benchmarks}.
            
        \subsubsection{Boson Sampling simulator} \label{sec:BS}
            Boson Sampling (BS) is a well-known linear optical quantum computing protocol introduced by Aaronson and Arkhipov~\cite{aaronson:2010}, which consists of sampling from the probability distribution of identical bosons scattered by a linear interferometer.
            Although in principle it applies to any bosonic system, its photonic version is the most natural one.
            While not universal, BS is strongly believed to be a classically computationally hard task. Traditionally its hardness has been shown in the \emph{collision-free} regime $d\gg n^2$ (where $d$ is the number of modes and $n$ is the number of particles) \cite{aaronson:2010}, but recently proof techniques have been extended to cover also a more feasible scenario in which $d$ scales linearly with $n$~\cite{Bouland2023}.

\begin{figure}
                \begin{centering}
                    \includegraphics[width=0.5\textwidth]{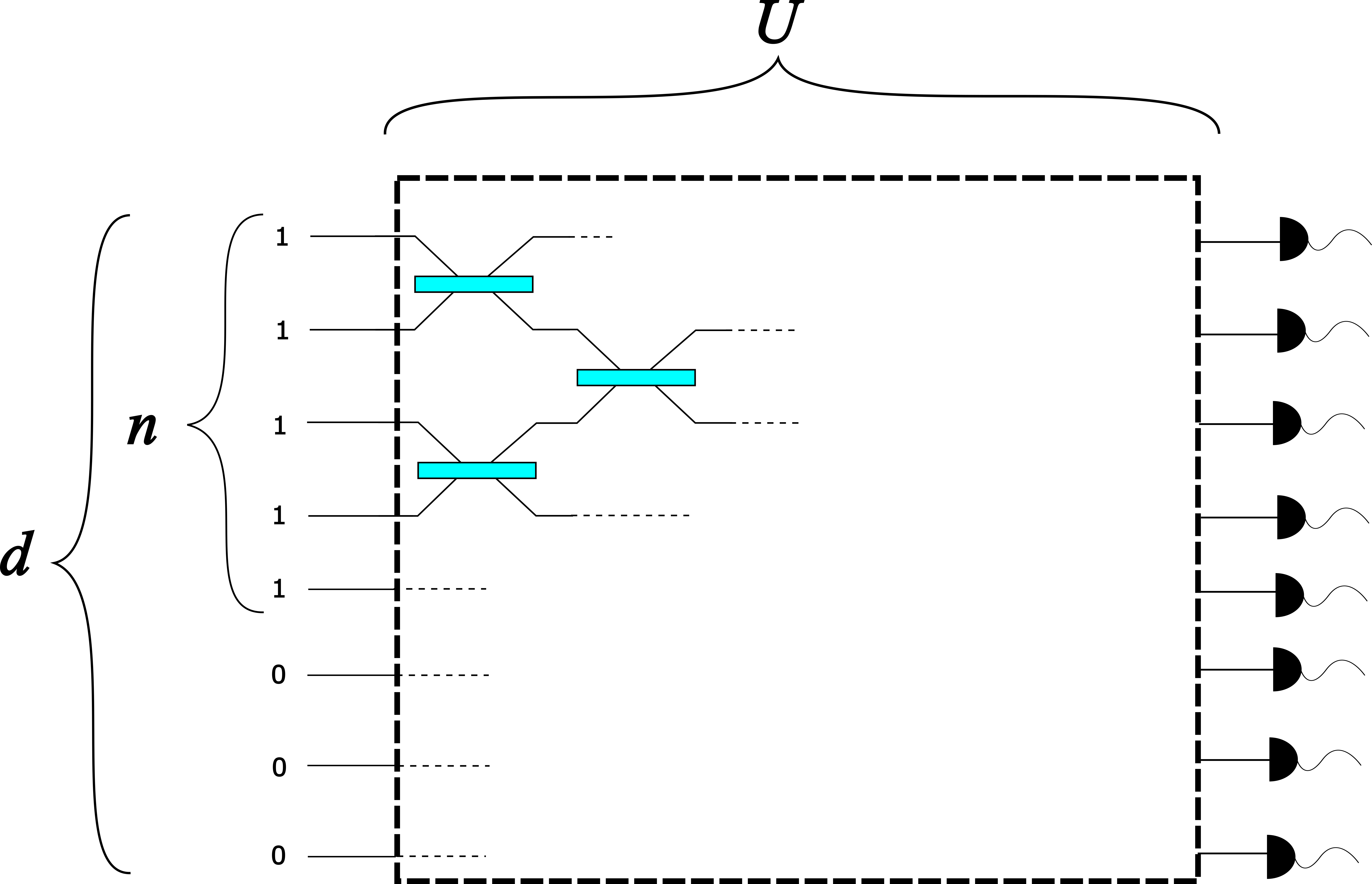}
                    \caption{Boson Sampling experiment setup.
                    The input state is an $n$-particle occupation number basis state, with typically only one or zero particles on each mode.  The particles are then scattered through a (particle number preserving) interferometer $U$, which can be decomposed into two-mode beamsplitters. The output state is then measured by number-resolving detectors.}
                    \label{fig:BS_Setup}
                \end{centering}
            \end{figure}

            In the standard BS scenario, depicted in Fig.~\ref{fig:BS_Setup}, the input state is a Fock basis state $\ket{\mathbf{s}}$ labeled by the occupation numbers $\mathbf{s} = (s_1, \ldots, s_d$) with $n=\sum_{i=1}^d s_i$ total number of photons. Typically the $s_i$'s are chosen to be $1$'s and $0$'s. The photons are then scattered in a generic passive linear optical interferometer described by a $d \times d$ unitary $U$. As a passive linear circuit preserves the particle number, the state vector is an element of the $\binom{n+d-1}{d-1}$ dimensional $n$-particle subspace during the entire evolution. Finally, a particle number measurement is performed on each mode, and the probability of the measurement outcome $\mathbf{t} = (t_1,t_2, \ldots, t_d)$ is given by the permanent introduced in Eq.~\eqref{eq:BS_prob} from Appendix~\ref{app:basics}.
            
            An example usage of the \lstinline{SamplingSimulator} class is shown in Code snippet~\ref{code:bosonsamplingexample}.
            The available gates are passive linear gates, and the only available measurement for this simulator is photon detection.

\begin{figure}[ht]
                \begin{centering}
                    \includegraphics[width = 0.25\textwidth]{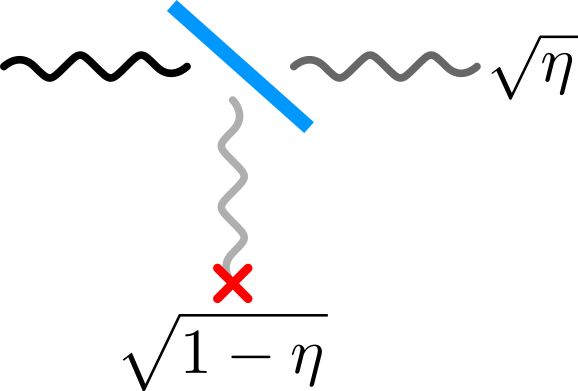}
                    \caption{Losses modeled by a beamsplitter. Assume that a photon in a given mode has a survival probability $\eta \in [0, 1]$. Such losses can theoretically be replaced by a beamsplitter transporting photons to an inaccessible mode, with a reflectance amplitude $\sqrt{1 - \eta}$.}
                    \label{fig:BS_LossModel}
                \end{centering}
            \end{figure}

            The effect of Gaussian losses on Boson Sampling can also be simulated in Piquasso. The simplest type is the pure-loss channel (with loss characterized by transmissivity $\eta$) on a mode, which can be modeled as an interaction of the mode with an environmental mode through a beamsplitter with transmittivity and reflectivity parameters $t=\sqrt{\eta}$ and $r=\sqrt{1-\eta}$, see Fig.~\ref{fig:BS_LossModel}.
            
            Given that the number of particles in the output states is generally smaller than in the inputs, one can no longer use a unitary (particle number preserving) matrix to model the interferometer. Instead, one has to use a matrix $A$ for which $AA^\dagger \leq \mathbb{I}$, where $\mathbb{I}$ denotes the identity (of proper size). Using SVD, $A$ can be rewritten as
            $A = V\operatorname{diag}(\mathbf{\mu})W$, where $V$ and $W$ are unitaries, and $\mathbf{\mu} = (\sqrt{\eta_1}, \sqrt{\eta_2}, ..., \sqrt{\eta_m})$ is the vector of singular values. The values $\eta_i \in [0, 1]$ can be interpreted as transmissions on the $i$-th mode \cite{MODB2018, MODB2020, Garcia2019}.

    \subsection{Extending Piquasso} \label{sec:extending}
        Piquasso is customizable by subclassing the abstract classes in the API as shown in Code snippet~\ref{code:customsimulator}. As an example, if the user only wants to customize a computation-heavy function, the new function could be specified by overriding the \lstinline{NumpyConnector} class, which could be defined in a \lstinline{Simulator} class as seen in Code snippet~\ref{code:customloophafnian}. Currently, supported built-in connectors are \lstinline{NumpyConnector} (default), \lstinline{TensorflowConnector} and \lstinline{JaxConnector}.

        One notable example of custom implementation is the PiquassoBoost plugin, delivering a capability beyond what conventional quantum computer emulator libraries offer. Specifically, it provides FPGA-based support for emulating Boson Sampling experiments~\cite{permanent_dfe}, a capability that sets it apart from other libraries.
        PiquassoBoost can be easily used along Piquasso as demonstrated by Code snippet~\ref{code:piquassoboost}. The source code of PiquassoBoost is made available at~\cite{piquassoboost}.

\section{Benchmarks}\label{sec:benchmarks}
    Piquasso aims to provide a collection of optimized algorithms to speed up the evaluation of specific computational tasks.
    During the development, several questions arose regarding performance. To answer these present and future questions, the algorithms are regularly examined and profiled using certain benchmarks and scripts.
    In this section, we present some of the current benchmarks in detail.

    \subsection{Gaussian Boson Sampling}
        The concept of the Gaussian Boson Sampling (GBS), i.e., the photon-detection measurement of a general Gaussian state, was first introduced in Refs.~\cite{PhysRevLett.119.170501, PhysRevA.100.032326}.
        Similarly to the conventional Boson Sampling, the classical resources needed to take a sample from a complex probability distribution of Gaussian photonic states tend to scale exponentially with the number of the involved photons making the problem classically intractable.
        The complexity of taking a sample from a non-displaced Gaussian state is characterized by the evaluation of a matrix function, called the hafnian
        (for details, see Refs.~\cite{PhysRevLett.119.170501, PhysRevResearch.2.023005,quesada2021quadratic}).
        The hafnian of a matrix can be considered as a generalization of the permanent:
        while the permanent enumerates the number of perfect matchings of a bipartite graph using its adjacency matrix, the hafnian yields the number of perfect matchings of a general graph.
        It was first introduced in Ref.~\cite{Caianiello1973} in a study of bosonic quantum field theory. For more details regarding
        Gaussian Boson Sampling see Appendix~\ref{app:particle_number}.
    
        Currently, the state-of-the-art calculation of the hafnian is the power trace algorithm with time complexity $O(n^3 2^{n/2})$ as described in Ref.~\cite{bjorklund2019faster},
        which can be further reduced in actual use cases when multiple photons occupy the same optical mode~\cite{Bulmer_2022}.
        These works also generalized the concept of hafnian for graphs including loops (i.e., generalized adjacency matrices containing nonzero diagonal elements) which enables the simulation of GBS with displaced Gaussian states as well~\cite{bjorklund2019faster}.
        The so-called loop hafnian differs from the hafnian by including corrections related to the nonzero diagonal matrix elements. The introduction of the loop hafnian also enabled a quadratic speed-up in the classical simulation of GBS~\cite{quesada2021quadratic}.

\begin{figure}
            \begin{centering}
                \includegraphics[width=0.5\textwidth]{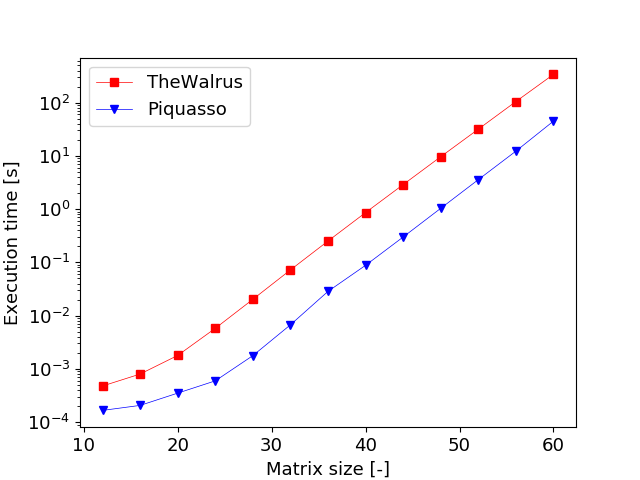}
                \caption{
                    Comparing the average repeated hafnian execution times of Piquasso (version 5.0.0) and TheWalrus (version 0.21.0) with increasing input matrix sizes. The repetitions were chosen to be $(2, \dots, 2)$ in each case. When the execution times are under 1 second, they are averaged over 100 runs, otherwise, a single run is executed. The benchmark was executed on a \emph{Intel Xeon E5-2650} processor platform. The script is available at~\cite{gaussian_boson_sampling_benchmark}. The script is available at~\cite{hafnian_benchmark}.
                }
                \label{fig:hafnian}
            \end{centering}
        \end{figure}

\begin{figure}
            \begin{centering}
                \includegraphics[width=0.5\textwidth]{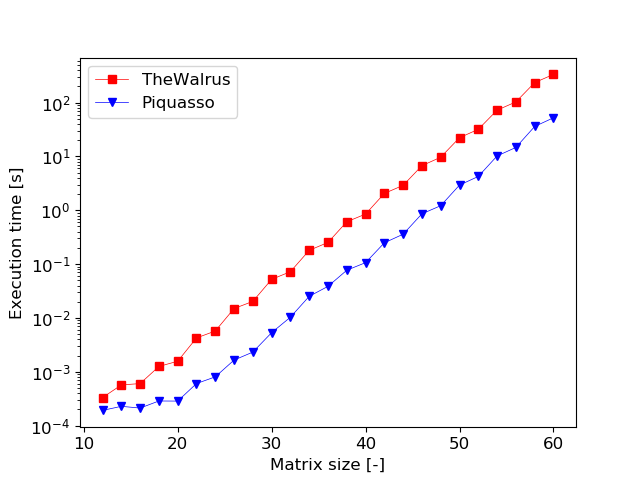}
                \caption{
                    Comparing the average repeated loop hafnian execution times of Piquasso (version 5.0.0) and TheWalrus (version 0.21.0) with increasing input matrix sizes. The repetitions were chosen to be $(2, \dots, 2)$ in each case. When the execution times are under 1 second, they are averaged over 100 runs, otherwise, a single run is executed. The benchmark was executed on a \emph{Intel Xeon E5-2650 processor} platform. The script is available at~\cite{loop_hafnian_benchmark}.
                }
                \label{fig:loop_hafnian}
            \end{centering}
        \end{figure}

\begin{figure}
            \centering
            \includegraphics[width=0.5\textwidth]{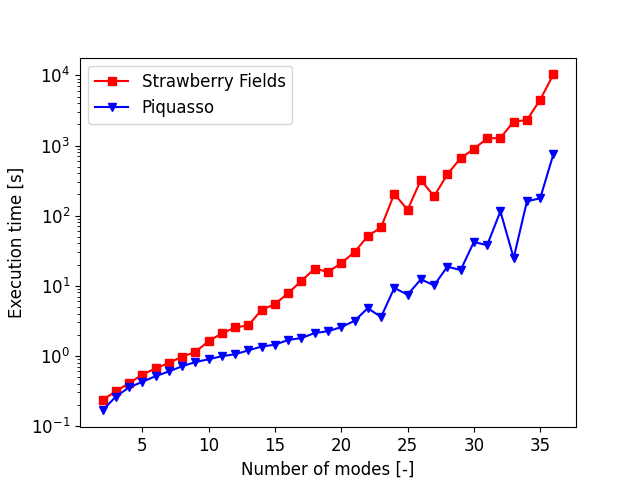}
            \caption{
                Performance benchmark of Gaussian Boson Sampling with Piquasso (version 5.0.0) and Strawberry Fields (version 0.23.0). The circuit was chosen to be a layer of squeezing gates with squeezing parameters $r = \operatorname{arcsinh}(1)$ consistently (to ensure that the average number of particles equal the number of modes), followed by an interferometer parametrized by a Haar-random unitary matrix. For each data point, 100 samples were taken.
                The benchmarks were executed on a \emph{AMD EPYC 7542 32-Core} processor platform. The script is available at~\cite{gaussian_boson_sampling_benchmark}.
            }
            \label{fig:gaussian_boson_sampling_benchmark}
        \end{figure}

        The implementation of the hafnian can be given in a Ryser-type (inclusion-exclusion) or a Glynn-type strategy, as in the implementation of the permanent function given in Refs.~\cite{ryser1963combinatorial,Glynn2013}.
        It turns out that the Glynn-type implementation is more precise than the Ryser formula. In the Ryser formulation, the inner addends are computed from submatrices of the input matrix, which can potentially lead to a wide range of addend magnitudes. In contrast, the Glynn variant derives its addends from matrices of consistent size, which might be the reason behind the significantly better numerical stability of this approach.
        The same conclusion holds on for the case of the loop hafnian and the permanent. We benchmarked the Glynn-type implementations of the hafnian and the loop hafnian between Piquasso and TheWalrus on Figure~\ref{fig:hafnian} and Figure~\ref{fig:loop_hafnian}, respectively.
        For completeness, we also compare the execution times of GBS between Piquasso and Strawberry Fields on Fig.~\ref{fig:gaussian_boson_sampling_benchmark}.
        In Piquasso, we implement the sampling algorithm as described in Ref.~\cite{quesada2021quadratic}, and we implement the same batching strategy as in Ref.~\cite{Bulmer_2022} for the loop hafnian, and we also use the halving of the Glynn-type iterations.

        The computational cost of evaluating the torontonian function, which was introduced for GBS with threshold detection \cite{PhysRevA.98.062322,li2020benchmarking}, was extensively studied in our previous work reported in Ref.~\cite{kaposi2022polynomial}. Here, we reported on a polynomial reduction on the computational reduction, that was based on the reuse of intermediate results. Quantum computing simulator softwares including Piquasso and Strawberry Fields already take advantage of this approach, providing the fastest way of computing the torontonian. By Ref.~\cite{PhysRevA.106.043712}, the method was further extended with loop corrections introducing the concept of loop torontonian, applicable for displaced Gaussian states. 

\begin{figure}
            \begin{centering}
                \includegraphics[width=0.5\textwidth]{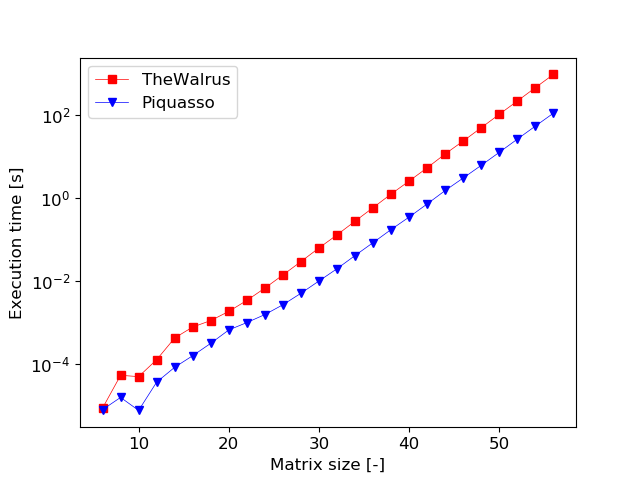}
                \caption{
                Comparing the average torontonian execution times of Piquasso (version 5.0.0) and TheWalrus (version 0.21.0) with increasing input matrix sizes. When the execution times are under 1 second, they are averaged over 100 runs, otherwise, a single run is executed. The benchmark was executed on a \emph{Intel Xeon E5-2650 processor} platform. The script is available at~\cite{torontonian_benchmark}.
                }
                \label{fig:torontonian}
            \end{centering}
        \end{figure}

\begin{figure}
            \begin{centering}
                \includegraphics[width=0.5\textwidth]{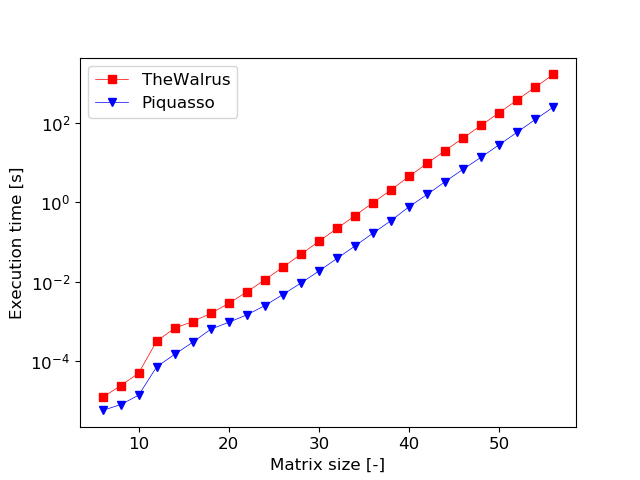}
                \caption{
                Comparing the average loop torontonian execution times of Piquasso (version 5.0.0) and TheWalrus (version 0.21.0) with increasing input matrix sizes. When the execution times are under 1 second, they are averaged over 100 runs, otherwise, a single run is executed. The benchmark was executed on a \emph{Intel Xeon E5-2650 processor} platform. The script is available at~\cite{loop_torontonian_benchmark}.
                }
                \label{fig:loop_torontonian}
            \end{centering}
        \end{figure}

        We show the performance benchmark results of the torontonian and the loop torontonian implementation on Figure~\ref{fig:torontonian} and Figure~\ref{fig:loop_torontonian}, respectively; comparing Piquasso (version 5.0.0) and TheWalrus (version 0.21.0).

    \subsection{Boson Sampling}
\begin{figure}
            \begin{centering}
                \includegraphics[width=0.5\textwidth]{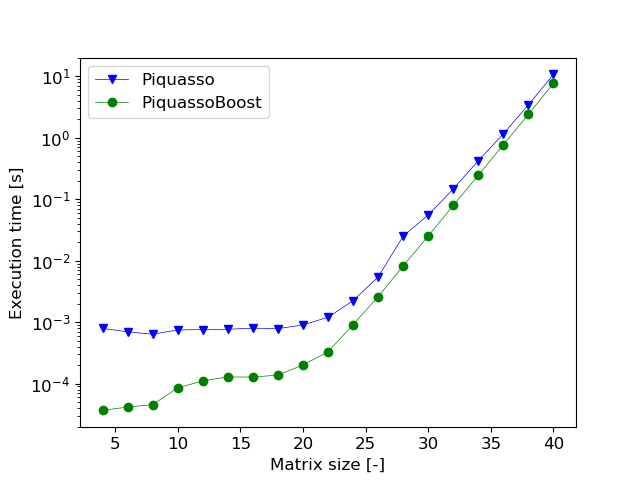}
                \caption{
                The average repeated permanent execution times of Piquasso (version 5.0.0) and PiquassoBoost with increasing input matrix sizes. The repetitions were chosen to be $(2, \dots, 2)$ in each case. The data points are calculated from an average of 100 runs. The benchmark was executed on a \emph{AMD EPYC 7542 32-Core} processor platform. The script used for benchmarking is available at~\cite{permanent_benchmark}.
                }
                \label{fig:permanent}
            \end{centering}
        \end{figure}

        As discussed in Sec.~\ref{sec:BS}, the permanent is the key mathematical function determining the performance of the Boson Sampling simulations.
        In our recent work, we have given a detailed analysis of state-of-the-art permanent calculation methods, including their performance and numerical precision~\cite{permanent_dfe}.
        We have shown that the main advantage of the Balasubramanian-Bax-Franklin-Glynn (BB/FG) formula~\cite{Glynn2013} over the Ryser variant~\cite{ryser1963combinatorial} lies in the precision of the calculated permanent value. The numerical experiments conducted with the MPFR multi-precision library~\cite{10.1145/1236463.1236468} revealed that Ryser's formula significantly lags behind the BB/FG method in terms of accuracy. 
        In order to increase the maximal number of photons in a simulation of an optical interferometer, we have extended the basic BB/FG approach to accommodate scenarios where the input matrix exhibits column or row multiplicities, representing multiple particles occupying a single photonic mode.
        This expansion is achieved by utilizing a generalized n-ary Gray code ordering for the outer sum in the BB/FG permanent formula. The digits in this code range from zero to the occupation number of individual optical modes. This generalization extends the applicability of the BB/FG formula, reducing computational complexity, as exemplified in our recent work on high-performance BS simulation~\cite{permanent_dfe}.
        Piquasso implements the proposed algorithm, providing better numerical accuracy in evaluating the permanent than alternative approaches~\cite{Chin2018,LUNDOW2022110990,clifford2020faster} based on the Ryser formula, while not losing computational performance. The perfornance benchmark of the Piquasso and PiquassoBoost implementation is shown in Fig.~\ref{fig:permanent}.

\begin{figure}
            \begin{centering}
                \includegraphics[width = 0.5\textwidth]{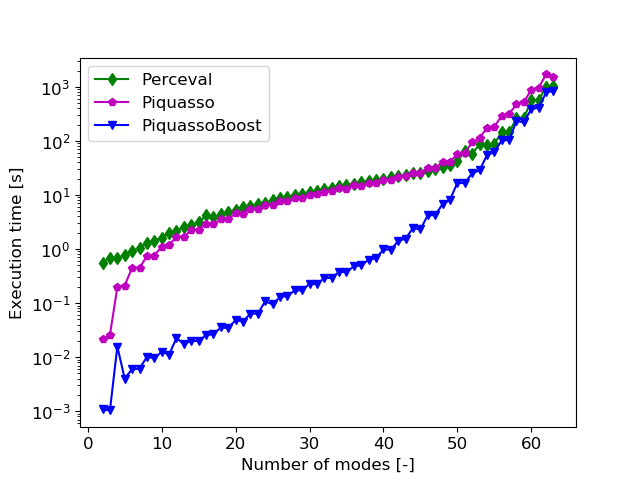}
                \caption{
                Performance benchmark of Boson Sampling with Piquasso (version 5.0.0), PiquassoBoost and Perceval (version 0.11.2). For the simulation, randomly generated Fock basis states were chosen as initial states, with number of particles half of the number of modes, and 100 samples were taken per datapoint under a runtime of 1s, otherwise 1 sample.
                The benchmarks were executed on a \emph{AMD EPYC 7542 32-Core} processor platform.
                The script used for benchmarking is available at~\cite{boson_sampling_benchmark}.
                }
                \label{fig:boson_sampling}
            \end{centering}
        \end{figure}

        In Piquasso, the permanent function is utilized
        for classically simulating ideal and lossy BS with Algorithm A in Ref.~\cite{clifford2020faster},
        which is the current state-of-the-art approach.
        This algorithm enables fast classical simulation of BS, as depicted in Figure~\ref{fig:boson_sampling}. Here, we compare Piquasso, PiquassoBoost, and Perceval as frameworks implementing efficient classical BS algorithms. This shows, that the execution times of Piquasso and Perceval are closely aligned, while PiquassoBoost has a significant speedup over both for smaller systems. For larger systems, the execution times of Perceval and PiquassoBoost are similar.
        Using PiquassoBoost, it is also possible to execute an enhanced algorithm with the help of data-flow engines, as described in Ref.~\cite{permanent_dfe}.

    \subsection{Fock-space based simulations}
        \label{sec:focksimulatorlosses}
        \begin{figure}
            \centering\centerline{\includegraphics[scale=0.50]{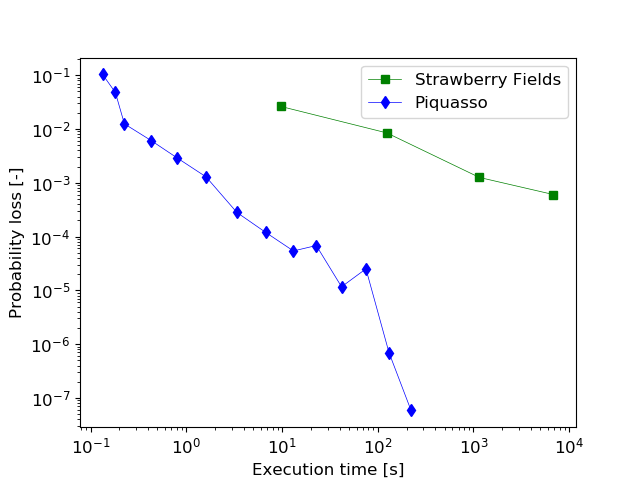}}
            \caption{
            Comparison of execution time and probability loss for Piquasso (version 5.0.0) and Strawberry Fields (version 0.23.0) Fock simulation. The benchmark was executed on $10$ modes for $4$ CVQNN layers, with variable global or local cutoff values. The range of global cutoff in Piquasso simulations was $[3, 16]$ and the range of local cutoff in Strawberry Fields simulations was $[3, 6]$. For lower execution times, the probability loss and execution time values are averaged over $10$ runs, while only $1$ calculation is executed when the execution time surpasses $10s$. The Kerr gate angles and the interferometer angles in the CVQNN layers were generated by uniform random angles drawn from the normal distribution $\mathcal{N}(0, 1)$, and the squeezing and displacement parameters were  drawn from $\mathcal{N}(0, 0.1)$.
            The benchmark was executed on a \emph{Intel Xeon E5-2650 processor} platform.
            The script is available at~\cite{probability_loss}.
            }
            \label{fig:probability_loss}
        \end{figure}
        
        As already discussed in Sec.~\ref{sec:focksimulator}, when using \lstinline{PureFockSimulator} and \lstinline{FockSimulator}, the truncation of the bosonic Fock space in Eq.~\eqref{eq:truncated_fock_space} may introduce loss of probability during the simulation. Storing a state vector in the truncated bosonic Fock space requires
        \begin{equation}
            \dim_{\mathbb{C}} \left ( \mathcal{F}^c_{B}(\C^d) \right ) = \binom{d + c - 1}{c - 1}
        \end{equation}
        complex numbers. This truncation amounts to a ``global'' cutoff, discarding any contribution to the state vector corresponding to a higher total particle number than the cutoff $c$. Note, that storing the state vector with a ``local'' cutoff instead would restrict the particle number for each mode individually. As a result, a pure state would practically be implemented as a tensor, which would require storing $c^d$ complex numbers, significantly higher than the value using global cutoff. Hence, for equal values of local cutoff and global cutoff, the state vector in the truncated Fock space scenario generally has fewer vector elements, possibly leading to a decrease in precision. On the bright side, one could argue that the contributions that are left out by using a global cutoff instead of a local one have small coefficients in most cases.
        Moreover, after the truncation using the global cutoff (for example on squeezed coherent states), the application of passive elements (e.g., Beamsplitter, Kerr) will not lead to further loss of probability during the simulation, while this is generally not true for the local cutoff.
        To illuminate this difference, the loss of probability has been calculated in two scenarios and illustrated in Fig.~\ref{fig:probability_loss}, where Piquasso implements a global cutoff, as opposed to the local cutoff implementation of Strawberry Fields. The benchmark was executed for $10$ modes and $4$ continuous-variable quantum neural network (CVQNN) layers, as described in~\cite{killoran2019continuous}.
        This benchmark shows that for any probability loss, the Piquasso Fock simulator implementing the global cutoff has a lower execution time than Strawberry Fields with a local cutoff.

    \subsubsection{Differentiability}
        For machine learning purposes, a fast calculation of the gradient corresponding to a photonic circuit has the utmost importance. To illustrate the ability of Piquasso to perform such tasks swiftly, we have performed benchmarks assessing the execution time of the gradient of a certain cost function.
        More specifically, the benchmarks introduced in this section are based on one learning step in a state learning algorithm, i.e., training a quantum circuit to fit a certain target state~\cite{arrazola2019machine}. The parameters of the quantum gates consisting of a single layer in the photonic circuit represent weights in the photonic neural network, and hence the gradient is calculated with respect to these parameters.

        Considering a target state $\ket{\psi_{\text{target}}}$, we can define a cost function based on the distance of the output state and the desired target state induced by the $2$-norm as
        \begin{equation} \label{eq:cost}
            J(\ket{\psi}) = \left|\left| \ket{\psi} - \ket{\psi_{\text{target}}} \right|\right|_2, 
        \end{equation}
        where $\ket{\psi}$ is the resulting state after executing the circuit, depending on the weights of the neural network.
        
\begin{figure}
            \centering
            \includegraphics[width=1.0\linewidth]{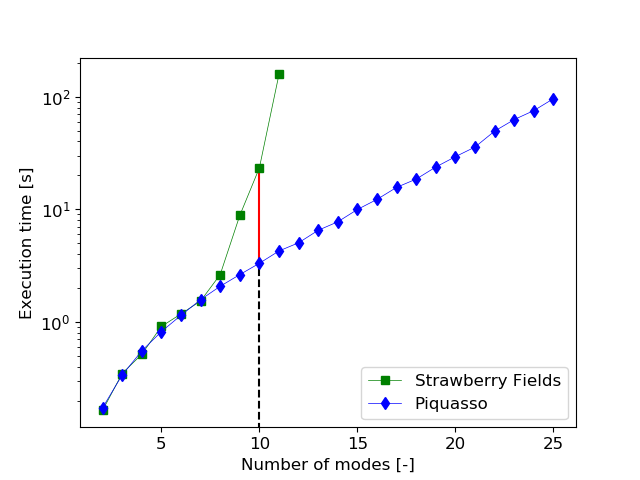}
            \caption{
            Comparison of gradient execution times of Piquasso  (version 5.0.0) using a global cutoff with the execution times of Strawberry Fields (version 0.23.0) using a local cutoff. The Piquasso implementation used a global cutoff of $6$, whereas the Strawberry Fields implementation uses a local cutoff of $4$, matching the probability loss $\sim 10^{-2}$ according to Figure~\ref{fig:probability_loss}. The circuit layout is the same as for Figure~\ref{fig:probability_loss}, and the weights were also chosen from the same distribution. The benchmarks were performed using \emph{AMD EPYC 7742P} architecture.
            The script is available at~\cite{cvqnn_perf}.
            }\label{fig:cvqnn_perf}
        \end{figure}

        The execution times of the cost function gradients are shown for Piquasso and Strawberry Fields in Fig.~\ref{fig:cvqnn_perf}. The benchmarks are executed
        with a global cutoff of $6$ for Piquasso and a local cutoff of $4$ for Strawberry Fields, matching the probability loss of $\sim 10^{-2}$ according to Figure~\ref{fig:probability_loss}.
        Note, that the probability loss comparison has been executed for $10$ modes, where the execution times match regardless of the framework. Surprisingly, the execution times approximately match until $8$ modes, which can be explained by the framework overhead of TensorFlow. However, this difference diminishes as we increase the number of modes. For example, as indicated by the red line in Figure~\ref{fig:probability_loss}, when the number of modes is $10$, a significant speedup of the global cutoff (Piquasso) implementation over the local cutoff implementation (Strawberry Fields) can be observed. In the Piquasso benchmark, the execution times increase exponentially for a given cutoff by increasing the number of modes, and the exponent appears to be linear. Meanwhile, in the Strawberry Fields benchmark, the execution times hit the exponential wall (in the logarithmic scale) by increasing the number of modes. In conclusion, the approaches using the local and global cutoff have a significant scaling difference in terms of the execution time.
        Moreover, due to the nature of the local cutoff (see Sec.~\ref{sec:focksimulatorlosses}) implemented in Strawberry Fields, its benchmark could not be executed for larger systems because of excessive memory usage.

\begin{figure}[ht!]
            \centering\centerline{\includegraphics[width=0.5\textwidth]{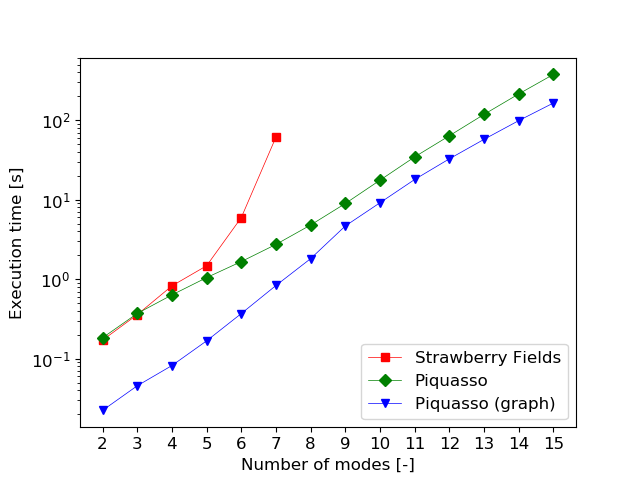}}
            \caption{
                Comparison of the execution times between the Piquasso \lstinline{PureFockSimulator} using \lstinline{TensorflowConnector} (version 5.0.0) with and without graph compilation and the Strawberry Fields Fock backend (version 0.23.0). During the calculation, the setup described by Code snippet~\ref{code:tffunction} was benchmarked with $4$ CVQNN layers and cutoff $10$ for both frameworks, the same setup as in Figure~\ref{fig:probability_loss}. When the execution times were under 1 second, the results were averaged over 100 runs, otherwise over 10 runs. We were forced to stop the calculation with Strawberry Fields due to excessive memory usage. The benchmarks were executed on a \emph{Intel Xeon E5-2650 processor} platform using 12 × 16GB of RAM.
                The script is available at~\cite{cvqnn_tensorflow_comparison_benchmark}
            }
            \label{fig:benchmark_tffunction}
        \end{figure}

        For repetitive tasks, it can be beneficial to compile Piquasso code using \lstinline{TensorflowConnector} into a callable TensorFlow graph, using \lstinline{tf.function}. This feature, after the initial trace-compilation is executed, provides a significant speedup in \lstinline{PureFockSimulator} by executing the optimized TensorFlow graph. A basic example is given by Code snippet~\ref{code:tffunction}. The graph execution via Piquasso is compared to Strawberry Fields with a benchmark on Figure~\ref{fig:benchmark_tffunction}, which shows a significant speedup of the compiled Piquasso code over Strawberry Fields.

\section{Web User Interface} \label{sec:ui}
    Piquasso offers an intuitive web-based interface that allows users to create photonic circuits seamlessly using an interactive drag-and-drop circuit composer, while also enabling them to collaborate with others and share their results effortlessly. Authentication is supported through social media accounts and unlocks additional features, including project saving and publication, as well as a streamlined collaboration with other users.

    \begin{figure*}[ht]
        \centering
        \includegraphics[width=\textwidth]{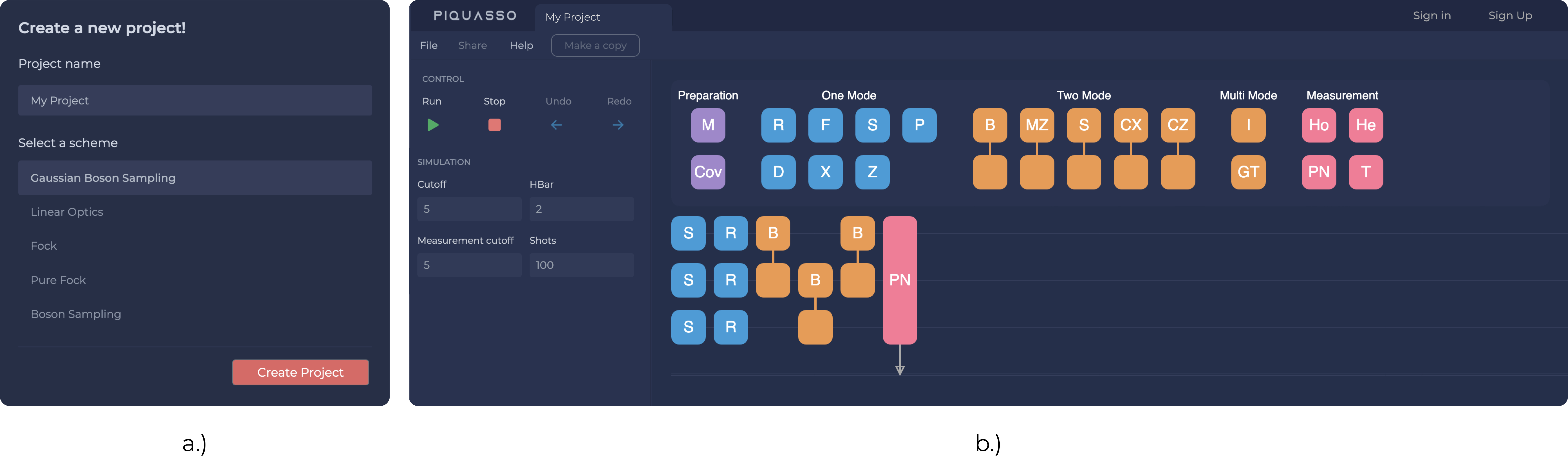}
        \caption{An overview of the Piquasso web interface. The user should (a) choose a supported backend scheme before creating a circuit, and then (b) use the interactive drag-and-drop composer to create the circuit from the relevant components of the chosen scheme. The depicted example is a Gaussian Boson Sampling circuit, the simulation results are visualized in Fig.~\ref{fig:ui_piquasso_results}.
        }
        \label{fig:ui_piquasso_overview}
    \end{figure*}

    \begin{figure*}[ht]
        \centering
        \includegraphics[width=\textwidth]{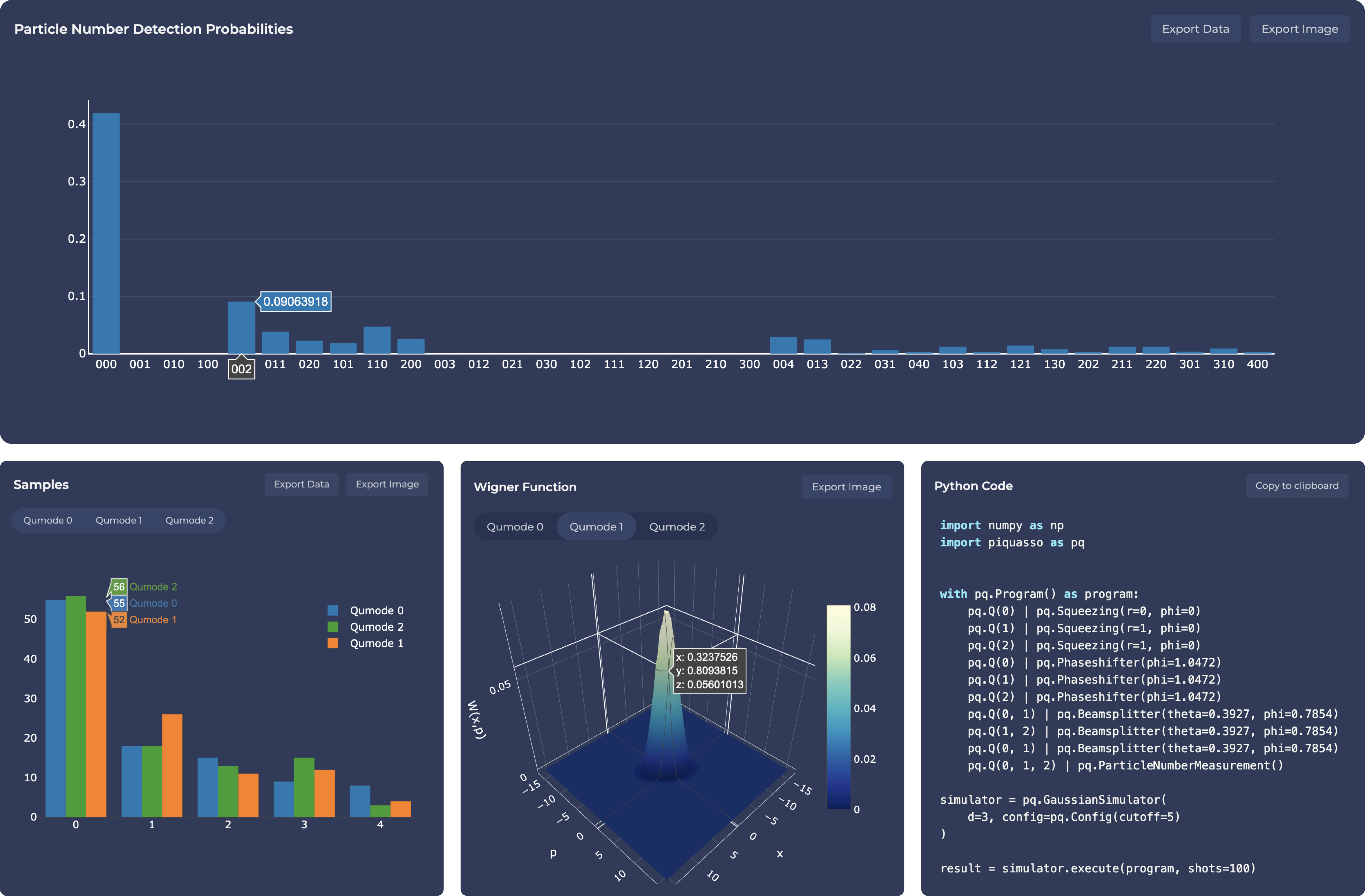}
        \caption{The simulated circuit results are visualized including the obtained measurement samples, the theoretical probability distribution, the Wigner function, and the corresponding Python code.}
        \label{fig:ui_piquasso_results}
    \end{figure*}

    \subsection{Starting a new project}
        When creating a new project, the user must begin by selecting one of the available backend schemes, which were described in Sec.~\ref{ssec:simulators}. Once a backend scheme has been selected, the Piquasso composer is launched. This tool provides an interactive drag-and-drop interface that enables the creation of photonic circuits with ease. The toolbar section of the interface displays the operations that are available for the selected scheme. These operations are grouped according to their respective types, such as 1-mode, 2-mode, multimode gates, measurements, and so on. To create a new circuit, the user simply needs to drag-and-drop the blocks that represent photonic operations onto the corresponding modes within the composer. The key components of the drag-and-drop interface are presented in Fig.~\ref{fig:ui_piquasso_overview}.
        
        Once the photonic circuit is finalized in the composer, the user can submit it through the control panel; the submitted job will be sent to the backend server for execution. The control panel offers several functionalities, such as starting or stopping a simulation and undoing or redoing previous actions. As soon as the submitted job is executed, the results are visualized on the dashboard, as shown in Fig.~\ref{fig:ui_piquasso_results}, and can be conveniently exported and downloaded in a variety of file formats, including \lstinline{PNG} and \lstinline{CSV}. 
        
        The project management menu provides users with an overview of their existing projects and of projects shared with them. Users can easily access previous projects and view detailed information about them including previously obtained results from past executions of the circuit. Furthermore, users can modify project settings such as the project name and collaborators directly from this menu.

    \subsection{Collaboration}
        One of our primary focus has been on enabling users to collaborate and share simulation data. To this end, we have implemented several features such as real-time data sharing, version control, and commenting functionality.

        The most prominent feature is the capability to add other Piquasso members directly to a project, enabling project members to collaborate in real-time on the same circuit and review the outcomes of submitted jobs. Depending on their role in the project, members can be granted different levels of access permissions, such as read-only access or full control over the circuit design. Write access allows users to freely modify the circuit, while read-only access allows users to view and execute the circuit yet forbidding any modifications. Project leaders have complete control over managing these permissions, as well as the ability to modify, rename, or share the project with others.
        
        Another powerful example of collaboration and data sharing is the publishing feature. Users can publish their circuits and quantum algorithms to the Piquasso community for others to see, run, and build upon. This feature opens up opportunities for users to share their work with the wider scientific community. Furthermore, these published circuits can be shared via a generated URL online, e.g., in publications or on social media platforms.

\section{Conclusion}
    We developed a photonic quantum computer simulator, named Piquasso, whose creation is driven by the increasing need for research in photonic quantum computing.
    The framework provides several specialized simulators tailored for specific computing schemes, along with a convenient and extendable Python interface, which have been explored in the current article.
    One such extension is the C/C++ plugin called PiquassoBoost which incorporates a high-performance engine into the Python framework, considerably aiding computational performance. Other extensions enable automatic differentiation of photonic circuits using popular machine learning frameworks such as TensorFlow and JAX.
    Finally, an ergonomic drag-and-drop web interface is made available at \href{https://piquasso.com}{piquasso.com}, which makes simulation data sharing and collaboration simple and accessible, empowering users to work together to push the boundaries of quantum computing and technological innovation.

\section{Acknowledgements}
    We would like to thank Zsófia Kallus, Gábor Németh and the Ericsson Research team for the inspiring discussions. We also thank the Wigner Scientific Computing Laboratory for their support.
    This research was supported by the Ministry of Culture and Innovation and the National Research, Development and Innovation Office through the Quantum Information National Laboratory of Hungary (Grant No. 2022-2.1.1-NL-2022-00004), the UNKP-22-5 New National Excellence Program, Grants No. K134437 and FK135220.
    ZZ also acknowledges the QuantERA II project HQCC-101017733.
    RP was supported by the Hungarian Academy of Sciences through the Bolyai J\'anos Stipendium (BO/00571/22/11).
    MO and TR acknowledge financial support from the Foundation for Polish Science via the TEAM-NET project (contract no. POIR.04.04.00-
    00-17C1/18-00).

\onecolumn
\newpage
\twocolumn

\bibliographystyle{quantum}
\bibliography{bibliography.bib}

\onecolumn
\newpage
\appendix

\renewcommand{\theequation}{\thesection.\arabic{equation}}

\section{Basics and notations} \label{app:basics}

    \subsection{Bosonic Fock space}
        The Hilbert space of a $d$-mode bosonic system 
        of $n$ particles is given by
        \begin{equation}
            \left(\C^{d}\right)^{\vee n} = \underbrace{\C^d \vee \C^d \vee \cdots \vee \C^d}_{\text{$n$ times}},
        \end{equation}
        where $\vee$ denotes the symmetrized tensor product. We also formally define the zero-particle Hilbert space $(\C^d)^{\vee 0}:= \C$, which corresponds to the vacuum. Given an orthonormal basis $\{ e_i \}_{i=1}^d$ on $\C^d$, we can directly construct an orthonormal basis on $\left(\C^{d}\right)^{\vee n}$ as
        \begin{equation}
            \left\{
                \left(
                    \tfrac{1}{n_1! \dots n_d!}
                \right)^{\frac {1}{2}}
                e_{i_1} {\vee} \dots {\vee} e_{i_n} : i_1 \leq {\dots} \leq i_n
            \right\},
        \end{equation}
        where $n_k$ denotes the repetition of the index $k$ in the index sequence $i_1, \ldots, i_n$ and $n_1 + \dots + n_d = n$. One could also denote the basis vectors more succinctly as
        \begin{equation} \label{eq:basis}
            \ket{n_1, \ldots, n_d} \coloneqq \left(
                    \tfrac{1}{n_1! \dots n_d!}
                \right)^{\frac {1}{2}}  
            e_{i_1} {\vee} \dots {\vee} e_{i_n}.
        \end{equation}
    
        Allowing for an arbitrary number of particles, the Hilbert space of a d-mode bosonic system is the bosonic Fock space given by the direct sum of all fixed particle Hilbert spaces
        \begin{equation} \label{eq:fockspace}
            \mathcal{F}_{B}(\C_d) =  \C \oplus \C^d \oplus \C^d \vee \C^d  \oplus \C^d \vee \C^d \vee \C^d \oplus \ldots
            = \bigoplus_{\lambda=0}^\infty  \, \left( \C^d \right)^{\vee \lambda}.
        \end{equation}
        An orthonormal basis of $ \mathcal{F}_{B}(\C_d)$ is given by the vectors $| n_1, \ldots, n_d \rangle$  with no restriction on the total particle number $n=n_1+n_2 \cdots +n_d$, this is called the {\it occupation number basis} of the Fock space.
        The annihilation and creation operators (jointly called ladder operators) are defined as
        \begin{align}
            \ann_j | n_1, ..., n_j, ..., n_d \rangle &= \sqrt{n_j} \, | n_1, ..., n_j {-} 1, ..., n_d \rangle, \nonumber \\
            \ann^\dagger_j | n_1, ..., n_j, ..., n_d \rangle &= \sqrt{n_j {+} 1} \, | n_1, ..., n_j {+} 1, ..., n_d \rangle. \nonumber
        \end{align}
        These operators satisfy the canonical commutation relations
        \begin{equation}
            [\ann_j, \ann_k] = [\ann^\dagger_j, \ann^\dagger_k] =0,  \quad
            [\ann_j, \ann^\dagger_k] = \delta_{j k} \mathbbm{1}.
        \end{equation}
    
    \subsection{Typical gates in photonic systems} \label{sec:typicalgates}
        Piquasso supports numerous photonic quantum gates, which we will review in this section. Piquasso can also export any \lstinline{Program} instance to Blackbird Quantum Assembly Language script~\cite{sf:2019}. To make the transition easier, we used the same parametrization of the gates as in Blackbird. Some of the most important ones are presented in this section.
    
        Except for Sec.~\ref{subsubsec:general}, the gates presented here are one- and two-mode gates, with their indices labeling on which modes they act non-trivially. The phase shift, the beamsplitter, the squeezing, and the displacement gates are Gaussian gates. More precisely, they map ladder operators to a linear combination of ladder operators (and also the identity in the case of the displacement gate), which we will also provide. The Kerr gate is a non-linear gate, however, it preserves the particle number similarly to the phaseshift and beamsplitter gates.
        
        \subsubsection{Phaseshift gate} \label{sec:rotation}    
            The phaseshift (or rotation) gate models the actual phase shifter optical element, which rotates the phase of a traveling electromagnetic wave. As a quantum gate, it acts on a state by rotating it in the canonical phase space. Additionally, the phaseshift gate is the only single-mode passive linear gate. 
            The unitary operator corresponding to the phaseshift gate is
            \begin{equation}
                R_j (\phi) = \exp \left (
                    i \phi \hat{n}_j
                \right ),
            \end{equation}
            where $\hat{n}_j \coloneqq \ann_j^\dagger \ann^{\phantom{\dagger}}_j$ and $\phi \in [0, 2\pi)$. The phaseshift gate is a passive linear optical element that transforms the ladder operators in the Heisenberg picture as follows:
            \begin{equation}
                R^\dagger_j (\phi)
                \begin{bmatrix}
                    \ann_j \\
                    \ann_j^\dagger
                \end{bmatrix}
                R_j (\phi)
                =
                \begin{bmatrix}
                    e^{i \phi } \ & 0 \\
                    0 & e^{- i \phi } 
                \end{bmatrix}
                \begin{bmatrix}
                    \ann_j \\
                    \ann_j^\dagger
                \end{bmatrix}.
            \end{equation}
            In Piquasso, a phaseshift gate can be implemented with \lstinline{pq.Phaseshifter}.

        \subsubsection{Beamsplitter gate} \label{sec:beamsplitter}
            A beamsplitter is an optical device that splits light into two parts and is an essential part of many optical experiments. The unitary operator corresponding to the beamsplitter gate is
            \begin{equation}
                B_{j k} (\theta, \phi) = \exp \left (
                \theta e^{i \phi} \ann^\dagger_j \ann_k
                    - \theta e^{- i \phi} \ann^\dagger_k \ann_j
                \right ),
            \end{equation}
            where $\theta, \phi \in [0, 2\pi)$. The beamsplitter gate transforms the ladder operators as
            \begin{equation}
                B^\dagger_{j k} (\theta, \phi) \begin{bmatrix}
                    \ann_j \\
                    \ann_k \\
                    \ann_j^\dagger \\
                    \ann_k^\dagger
                \end{bmatrix}
                B_{j k} (\theta, \phi)
                \\=
                \begin{bmatrix}
                    t  & -r^* &    & \\
                    r & t   &    & \\
                       &     & t  & -r \\
                       &     & r^* & t
                \end{bmatrix}
                \begin{bmatrix}
                    \ann_j \\
                    \ann_k \\
                    \ann_j^\dagger \\
                    \ann_k^\dagger
                \end{bmatrix},
            \end{equation}
            where $t = \cos(\theta)$ and $r = e^{i \phi} \sin(\theta)$.
            In Piquasso, a beamsplitter gate can be implemented with \lstinline{pq.Beamsplitter}.

        \subsubsection{Squeezing gate}
            Considering a vacuum state as the initial state, the squeezing gate produces a state that is ``squeezed'' in the phase space along a certain direction.
            The squeezing gate is the prototypical example of an active linear gate.
            The unitary operator corresponding to the squeezing gate is
            \begin{equation}
                S_{j} (z) = \exp \left (
                    \frac{1}{2}\left (z^* \ann_j^2 - z \ann_j^{\dagger 2} \right)
                \right ),
            \end{equation}
            where $z \in \mathbb{C}$.
            The squeezing gate transforms the ladder operators as
            \begin{equation}
                S_{j}^\dagger(z) \begin{bmatrix}
                    \ann_j \\
                    \ann_j^\dagger
                \end{bmatrix}
                S_{j}(z)
                \\=
                \begin{bmatrix}
                    \cosh r & - e^{i \phi} \sinh r \\
                    - e^{- i \phi} \sinh r & \cosh r
                \end{bmatrix}
                \begin{bmatrix}
                    \ann_j \\
                    \ann_j^\dagger
                \end{bmatrix},
            \end{equation}
            where $z = r e^{i\phi}$.
            In Piquasso, a squeezing gate can be implemented with \lstinline{pq.Squeezing}.

        \subsubsection{Displacement gate}
            A displacement gate ``displaces'' the quantum state in the phase space. Starting from vacuum as an initial state, the resulting states after applying a displacement gate are called coherent states.
            The unitary operator corresponding to the displacement gate is
            \begin{equation}
                D_j(\alpha) = \exp \left ( \alpha \ann_j^\dagger -  \alpha^* \ann_j \right ),
            \end{equation}
            where $\alpha \in \mathbb{C}$.
            The displacement gate transforms the ladder operators as
            \begin{equation}
                D^\dagger_j(\alpha)
                \begin{bmatrix}
                    \ann_j \\
                    \ann_j^\dagger
                \end{bmatrix}
                D_j(\alpha)
                =
                \begin{bmatrix}
                    \ann_j + \alpha \mathbbm{1} \\
                    \ann_j^\dagger + \alpha^* \mathbbm{1}
                \end{bmatrix}.
            \end{equation}
            In Piquasso, a displacement gate can be implemented with \lstinline{pq.Displacement}.

        \subsubsection{Kerr gate}
            The Kerr gate is the most trivial example of a passive non-linear gate. The definition of the Kerr gate is
            \begin{equation}
                K_j (\kappa) = \exp \left (
                    i \kappa \hat{n}^2_j
                \right ),
            \end{equation}
            where $\kappa \in [0, 2\pi)$. The Kerr gate transforms the ladder operators as
            \begin{align}
                K^\dagger_j(\kappa) \ann_j K_j(\kappa) &= \ann_j \exp\left(i \kappa \left( 2 \hat{n}_j - \mathbbm{1} \right)\right), \nonumber \\
                K^\dagger_j(\kappa) \ann_j^\dagger K_j(\kappa) &= \ann_j^\dagger \exp\left(-i \kappa \left( 2 \hat{n}_j - \mathbbm{1} \right)\right).
            \end{align}
            In Piquasso, a Kerr gate can be implemented with \lstinline{pq.Kerr}.

        \subsubsection{General Gaussian gates} \label{subsubsec:general}
            Gaussian unitaries can be characterized by Hamiltonian operators which contain only quadratic and linear terms. Concretely, one can write the Hamiltonian as
            \begin{align}
                H  &= \sum_{j, k = 1}^d A_{j k} \ann_j^\dagger \ann_k + B_{j k} \ann_j^\dagger \ann^\dagger_k + \sum_{j = 1}^d \beta_j \ann_j + h.c. \nonumber \\
                   &= \xi^\dagger \textbf{H} \xi + \alpha \xi ,
            \end{align}
            with
            \begin{align}
                \textbf{H} &= \begin{bmatrix}
                    A & B     \\
                    B^* & A^* \\
                \end{bmatrix}, \\
                \xi &= \left [
                    \ann_1, 
                    \dots,
                    \ann_d,
                    \ann_1^\dagger,
                    \dots,
                    \ann_d^\dagger
                \right]^T,
            \end{align}
            where $A = A^\dagger$, $B = B^T$, $ \alpha = \left[ \beta, \beta^* \right]$, and $ \beta \in \mathbb{C}^d$.
            When $ \beta = 0_{d} $ and $B=0_{d \times d}$, the Hamiltonian $H$ corresponds to passive Gaussian gates, e.g., beamsplitters and phaseshifters. These are Gaussian gates that preserve the particle number. Generally, the quadratic part of the Hamiltonian can be split into passive and active parts, while the linear terms correspond to displacements. In particular, for  $A=B = 0_{d \times d}$ and general $ \beta \in \mathbb{C}^{d} $ the Hamiltonian $H$ describes pure displacement gates. 
            
            Under a Gaussian transformation, Gaussian states are mapped to Gaussian states. The symplectic representation of unitary evolution can be described by
            \begin{equation} \label{eq:symplectic_unitary}
                U^\dagger \xi U = S_{(c)} \xi + \alpha,
            \end{equation}
            where $\alpha \in \mathbb{C}^{2 d}$, $U = \exp \left( - i \left ( \xi^\dagger \mathbf{H} \xi + \alpha \xi \right ) \right)$ and $  S_{(c)}  \in \operatorname{Sp}(2 d) $ is a symplectic matrix. Then $ S_{(c)} $ can be calculated by
            \begin{equation} \label{eq:symplectic}
                S_{(c)} = e^{- i K \mathbf{H}}\,, \quad \text{where} \quad
                K = \begin{bmatrix}
                    \mathbbm{1} &  \\
                      & - \mathbbm{1}
                \end{bmatrix}.
            \end{equation}
            In Piquasso, a general Gaussian tranformation can be implemented with \lstinline{pq.GaussianTransform}.
            
    \subsection{States and their evolution} \label{sec:states}
        In the simulation of photonic quantum computation, states can 
        be represented in multiple ways. In this subsection, we will provide a short overview of the most important representations. We also describe the action of the most important gates in the different representations.
        
        \subsubsection{Generic states in the occupation number representation}
            A pure bosonic state is an element of the multimode bosonic Fock space defined in Eq.~\eqref{eq:fockspace}. Generally, it can be expanded in the canonical basis of this vector space, i.e., in the occupation number basis as
            \begin{equation}
                | \psi \rangle = \sum_{\mathbf{n}  \in \mathbb{N}^d} c_{\mathbf{n}} | \mathbf{n} \rangle,
            \end{equation}
            where $\langle \psi | \psi \rangle = \sum_{ n \in \mathbb{N}^d} | c_{n} |^2 = 1$.
            Similarly, a general mixed state can be written as a density operator
            \begin{equation}
                \rho = \sum_{\mathbf{n}, \mathbf{m} \in \mathbb{N}^d} c_{ \mathbf{n}, \mathbf{m}} | \mathbf{n} \rangle \langle \mathbf{m} |,
            \end{equation}
            where $\operatorname{Tr} \rho = 1$ and $\rho$ is positive semidefinite.
    
            Given a quantum gate from Sec.~\ref{sec:typicalgates}, the transformation of the ladder operators in the Heisenberg picture relates to the evolution of the quantum states in the Fock representation by Eq.~\eqref{eq:symplectic_unitary}. 
            As an example, by applying a rotation gate from Sec.~\ref{sec:rotation} to an occupation number state, one gets
            \begin{equation}
                R_j (\phi) | n_1 \dots n_j \dots n_d \rangle \\= e^{i n_j \phi} | n_1 \dots n_j \dots n_d \rangle.
            \end{equation}
            Similarly, for a beamsplitter gate, one can write
            \begin{equation}
                B_{j k} (\theta, \phi)
                | n, m \rangle =
                \\
                \sum_{k=0}^{n}
                \sum_{l=0}^{m}
                c_{n, m}^{k, l}(r, t)
                |
                    n {-} k {+} l,
                    m {+} k {-} l
                \rangle,
            \end{equation}
            where $c_{n, m}^{k, l}(r,t)= \binom{n}{k}
                \binom{m}{l}
                t^{n + m - k - l}
                (- r^*)^{l}
                r^{k}$.
    
            When an active gate is applied to a fixed particle-number state, the resulting state will have terms in different particle-number sectors.
            As an illustration, the squeezing gate is applied to the vacuum as
            \begin{equation}
                S_j (z) | 0_1 \dots 0_j \dots 0_d \rangle = 
                \frac{1}{\sqrt{\cosh r}} \, \sum_{k_j=0}^\infty \left(
                    - e^{i\phi} \tanh r
                \right)^{k_j}
                \frac{\sqrt{(2k_j)!}}{2^{k_j} k_j!} | 0_1 \dots 2k_j \dots 0_d \rangle,
            \end{equation}
            where $r = |z|$ and $\phi = \arg(z)$.
        
            Finally, the action of the non-linear Kerr gate applied to an occupation number basis state is given as
            \begin{equation}
                K_j (\phi) | n_1 \dots n_j \dots n_d \rangle
                \\= \exp \left( i \xi n_j^2 \right) | n_1 \dots n_j \dots n_d \rangle.
            \end{equation}
            Similar results can be obtained when applying the above considerations for density operators.
            For such calculations, one could use \lstinline{pq.PureFockSimulator} or \lstinline{pq.FockSimulator} in Piquasso.
    
        \subsubsection{Gaussian states}
            Gaussian states on $d$ modes can be characterized by their mean vector and covariance matrix $(\mu, \sigma)$ defined by
            \begin{align} \label{eq:meancov}
            \begin{split}
                \mu_i &= \operatorname{Tr} \left [ \rho R_i  \right ], \\
                \sigma_{ij} &= \operatorname{Tr} \left [ \rho \{
                    R_i - \mu_i, R_j - \mu_j
                \} \right ], \qquad ( i,j = 1, \dots, 2 d )
            \end{split}
            \end{align}
            where $\{A, B\} = AB + BA$ is the anticommutator and $R$ is a vector of operators
            \begin{equation} \label{eq:xporder}
                R = (
                    \hx_1,
                    \dots,
                    \hx_d,
                    \hp_1,
                    \dots,
                    \hp_d
                )^T.
            \end{equation}
            
            The evolution of the quantum state in terms of the mean vector and the covariance matrix can be given in terms of the evolution of the ladder operators in the Heisenberg picture as
            \begin{align}
                \mu &\mapsto S \mu \nonumber, \\
                \sigma &\mapsto S \sigma S^T,
            \end{align}
            where $S$ is the symplectic matrix corresponding to the quantum gate in the canonical coordinate representation according to Eq.~\eqref{eq:xporder}.
            For simulating Gaussian states, one could use \lstinline{pq.GaussianSimulator} in Piquasso.
    
    \subsection{Typical measurements}\label{app:typical_measurements}
    
        \subsubsection{Homodyne and heterodyne measurements}
        
            Homodyne and heterodyne measurements correspond to the usual detection schemes from optical setups. These measurements have the common property of preserving the Gaussian character of the state after measurement. Homodyne measurement corresponds to the measurement of the operator
            \begin{equation}
                \hat{x}_{\phi} = \cos (\phi) \hat{x} + \sin (\phi) \hat{p}
            \end{equation}
            with outcome probability density given by
            \begin{equation}
                p(x_{\phi}) = \langle x_{\phi} | \rho | x_{\phi} \rangle,
            \end{equation}
            where $x_{\phi}$ correspond to the eigenvalues of $\hat{x}_{\phi}$. On the other hand, using heterodyne measurement, the probability density is given by
            \begin{equation}
                p(x_{\phi}) = \frac{1}{\pi} \operatorname{Tr} \left [
                    \rho | \alpha \rangle \langle \alpha |
                \right ].
            \end{equation}
            In optical setups, the heterodyne measurement is performed by mixing the state $\rho$ with a vacuum state $\ket{0}$, then subtracting the detected intensities of the two outputs. The mixing is performed with a 50:50 beamsplitter.
            Homodyne and heterodyne measurements can be implemented using \lstinline{pq.HomodyneMeasurement} and \lstinline{pq.HeterodyneMeasurement} in Piquasso.
        
        \subsubsection{Particle number measurement} \label{app:particle_number}
            Particle number measurement or photon detection is a non-Gaussian projective measurement that is performed via number-resolving detectors. The probability of detecting particle numbers $\mathbf{n} = (n_1, n_2, \ldots, n_d)$ is given by
            \begin{equation}
                p(\mathbf{n}) = \operatorname{Tr} \left [ \rho | \mathbf{n} \rangle \langle \mathbf{n} | \right ].
            \end{equation}
            The samples are non-negative integer values corresponding to the detected photon number.
            
            When the state is a non-displaced Gaussian state, the probability of an output $\mathbf{t} = (t_1, \dots, t_d)$ during particle number measurement is given by
            \begin{equation}
                p(\mathbf{t}) = \frac{1}{\sqrt{\operatorname{det}Q}} \frac{
                    \operatorname{haf} \left( A_{\mathbf{t}} \right)
                }{t_1! \dots t_d!},
            \end{equation}
            where $A_{\mathbf{t}}$ is the block reduction of the matrix $A$ by the vector $\mathbf{t}$ and
            \begin{align}
                A &= \begin{bmatrix}
                      & \mathbbm{1} \\
                    \mathbbm{1} & \\
                \end{bmatrix}
                \left(\mathbbm{1} - Q^{-1}\right), \nonumber \\
                Q &= \Sigma + \frac{1}{2} \mathbbm{1}, \label{eq:husimi} \\
                \Sigma &= W \sigma W^\dagger, \nonumber \\
                W &= \frac{1}{\sqrt{2}} \begin{bmatrix}
                    \mathbbm{1} & i \mathbbm{1} \\
                    \mathbbm{1} & -i \mathbbm{1} \\
                \end{bmatrix} \nonumber
            \end{align}
            following Hamilton et. al.~\cite{PhysRevLett.119.170501}. 
            
            Furthermore, the hafnian of a matrix is defined by
            \begin{equation} \label{eq:haf_def}
                \operatorname{haf}(A) = \sum_{M \in \operatorname{PMP}(n)} \prod_{(i,j) \in M} A_{i, j},
            \end{equation}
            where $\operatorname{PMP}(n)$ is the set of perfect matching permutations of $n$ even elements, such that $\sigma(2i-1)<\sigma(2i)$ and $\sigma(2i-1)<\sigma(2i+1)$, where $\sigma: [n] \to [n]$ is some permutation.

            When the Gaussian state is displaced with complex displacement vector $\bm{\alpha}$, the detection probability of an output $\mathbf{t}$ is given by
            \begin{equation}
                p(\mathbf{t}) = \frac{\exp(-\frac{1}{2}\bm{\alpha}^\dagger\Sigma^{-1}\bm{\alpha} )}{\sqrt{\operatorname{det}Q}} \frac{
                \operatorname{lhaf} \left( \mathrm{filldiag}(A_{\mathbf{t}}, \bm{\gamma}_{\mathbf{t}}) \right)
            }{t_1! \dots t_d!},
            \end{equation}
            where $\bm{\gamma} = \bm{\alpha}^\dagger \Sigma^{-1}$ and $\mathrm{filldiag}$ puts the repeated vector $\bm{\gamma}_\mathbf{t}$ into the diagonals of $A_{\mathbf{t}}$. Moreover, the $\mathrm{lhaf}$ is the loop hafnian of a matrix $A \in \mathbb{C}^{n\times n}$ is defined by~\cite{Quesada_2019}
            \begin{equation} \label{eq:loop_hafnian}
                \operatorname{lhaf}(A) = \sum_{M \in \operatorname{SPM}(n)} \prod_{(i,j) \in M} A_{i, j},
            \end{equation}
            where $\operatorname{SPM}(n)$ is the set of single-pair matchings of $n$ elements, which is analoguous to $\operatorname{PMP}(n)$, but loops are also permitted. Particle number measurement using Gaussian states is also called Gaussian Boson Sampling.
            
            When the state is an occupation number state with $\mathbf{s} = (s_1, \dots, s_d)$ initial particles, the probability of resulting in $\mathbf{t} = (t_1, \dots, t_d)$ particles after unitary circuit is~\cite{aaronson:2010}
            \begin{equation} \label{eq:BS_prob}
                p(\mathbf{s}, \mathbf{t})=
                \frac{
                    |\operatorname{per}
                    \left ( U_{\mathbf{s}, \mathbf{t}} \right )|^2
                }{
                    t_1! \dots t_d! s_1! \dots s_d!
                },
            \end{equation}
            where $U_{\mathbf{s}, \mathbf{t}}$ represents the set of unitary gates $U$ corresponding to the circuit, and the permanent $\operatorname{per}$ of a matrix is defined via
            \begin{equation} \label{eq:per_def}
                \operatorname{per}(A) = \sum_{\sigma \in S_n}  \prod_{i = 1}^n A_{\sigma(i), i},
            \end{equation}
            where $A \in \mathbb{C}^{n \times n}$.
            Particle number measurements can be implemented using \lstinline{pq.ParticleNumberMeasurement} in Piquasso.
    
        \subsubsection{Threshold measurement}\label{app:threshold}
            Threshold measurement or threshold detection is very similar to particle number measurement, but it only results in samples containing $0$ or $1$, where $0$ corresponds to no photon being detected, and $1$ corresponds to the detection of at least one photon.
            
            Note that threshold measurement is only supported in the Gaussian simulator. When the state is a non-displaced Gaussian state, the probability of an output $\mathbf{t} = (t_1, \dots, t_d) \in \{0, 1\}^d$ during threshold detection is given by~\cite{PhysRevA.98.062322}
            \begin{equation}
                p(\mathbf{t}) = \frac{1}{\sqrt{\operatorname{det}Q}} \frac{
                    \operatorname{tor} \left( O_{\mathbf{t}} \right)
                }{t_1! \dots t_d!},
            \end{equation}
            where
            \begin{equation}
                O = \mathbbm{1} - Q^{-1},
            \end{equation}
            $Q$ is given by Eq.~\eqref{eq:husimi} and the torontonian $\operatorname{tor}$ is
            defined as
            \begin{equation} \label{eq:tor_def}
                 \operatorname{tor}(A) = \sum\limits_{\mathbf{z} \in P_n}  \frac{(-1)^{n/2-|\mathbf{z}|}}{\sqrt{|\operatorname{det}(\mathbbm{1}-A_{\mathbf{z}})|}},
            \end{equation}
            where $P_n$ is the power set of $1,2,\dots, n/2$.
            When the Gaussian state is displaced with complex displacement vector $\bm{\alpha}$, the probability distribution is modified as~\cite{PhysRevA.106.043712}
            \begin{equation}
                p(\mathbf{t}) = \frac{\exp(-\frac{1}{2}\bm{\alpha}^\dagger\Sigma^{-1}\bm{\alpha} )}{\sqrt{\operatorname{det}Q}}
                \operatorname{ltor}( O_{\mathbf{t}}, \bm{\gamma}_{\mathbf{t}}),
            \end{equation}
            where $\bm{\gamma} = (\Sigma^{-1} \bm{\alpha})^* $ and $\operatorname{ltor}$ is the loop torontonian defined by
            \begin{equation}
                \operatorname{ltor}(A, \bm{\gamma})
                =
                \sum\limits_{\mathbf{z} \in P_n}  \frac{(-1)^{n/2-|\mathbf{z}|}}{\sqrt{|\operatorname{det}(\mathbbm{1}-A_{\mathbf{z}})|}} \exp(
                    \frac{1}{2} \bm{\gamma}^T (\mathbbm{1} - A_{\mathbf{z}})^{-1} \bm{\gamma}^*
                ).
            \end{equation}
            Threshold measurements can be implemented using \lstinline{pq.ThresholdMeasurement} in Piquasso.

\newpage
\section{Code snippets}
    The code examples in this Appendix are written for Piquasso version 5.0.0.

    \begin{figure*}[htbp]
\begin{lstlisting}[
            language=Python,
            caption={ Example usage of \lstinline{PureFockSimulator}, where the initial state is specified as $\frac{1}{\sqrt{2}} (\ket{1 1 1} + \ket{0 1 1})$. For simulations where the states are represented in the Fock space, a Fock space \textit{cutoff} needs to be specified in a \lstinline{pq.Config} instance, which is then passed to \lstinline{PureFockSimulator} as a constructor parameter. The \lstinline{FockSimulator} can be parametrized similarly.},
            label={code:purefockexample}
        ]
import numpy as np
import piquasso as pq

with pq.Program() as program: 
    pq.Q(all) | pq.StateVector((1, 1, 1)) / np.sqrt(2)
    pq.Q(all) | pq.StateVector((0, 1, 1)) / np.sqrt(2)

    pq.Q(1,2) | pq.CrossKerr(xi=np.pi / 4)

    pq.Q(all) | pq.ParticleNumberMeasurement()
 
config = pq.Config(cutoff=4)
simulator = pq.PureFockSimulator(d=3, config=config)
result = simulator.execute(program, shots=1000)
print(result.samples)  # [(0, 1, 1), (0, 1, 1), ...
\end{lstlisting}
    \end{figure*}

    \begin{figure*}[htbp]
\begin{lstlisting}[
            language=Python,
            caption={Example usage of \lstinline{GaussianSimulator}. This simulator enables fast simulation of Gaussian states, which are states that can be constructed via linear optical gates from the vacuum (or from single-mode thermal states in the mixed case). In this example, these linear gates are the squeezing, the displacement, and the beamsplitter gates. The program definition is concluded with a photon detection on all modes. Finally, the program is executed via the simulator with 100 shots.},
            label={code:gaussianexample}
        ]
import numpy as np
import piquasso as pq

simulator = pq.GaussianSimulator(d=5)

with pq.Program() as program:
    pq.Q(all) | pq.Vacuum()

    for i in range(5):
        pq.Q(i) | pq.Squeezing(r=0.1) | pq.Displacement(r=1.0)

    pq.Q(0,1) | pq.Beamsplitter(theta=np.pi / 3)
    pq.Q(2,3) | pq.Beamsplitter(theta=np.pi / 4)
    pq.Q(3,4) | pq.Beamsplitter(theta=np.pi / 5)

    pq.Q(all) | pq.ParticleNumberMeasurement()

result = simulator.execute(program, shots=1000)
print(result.samples)
# [(0, 1, 1, 2, 2), (1, 3, 0, 0, 1), (0, 1, 0, 0, 3),...
\end{lstlisting}
    \end{figure*}

    \begin{figure*}[htbp]
\begin{lstlisting}[
            language=Python,
            caption={Example usage of \lstinline{SamplingSimulator}, which is specifically tailored for the Boson Sampling algorithm. Compared to \lstinline{PureFockSimulator}, this simulator is restricted in a way that only \lstinline{StateVector} can be used as preparation, only \lstinline{ParticleNumberMeasurement} as measurement, and only passive linear elements as gates.},
            label={code:bosonsamplingexample}
        ]
import numpy as np
import piquasso as pq

with pq.Program() as program:
    pq.Q(all) | pq.StateVector((1, 2, 0, 2, 1))
    
    pq.Q(1,2) | pq.Beamsplitter(theta=np.pi / 3)
    pq.Q(2,3) | pq.Beamsplitter(theta=np.pi / 4)
    pq.Q(0,1) | pq.Beamsplitter(theta=np.pi / 5)
    pq.Q(3,4) | pq.Beamsplitter(theta=np.pi / 6)

    pq.Q(all) | pq.ParticleNumberMeasurement()

simulator = pq.SamplingSimulator(d=5)
result = simulator.execute(program, shots=1000)
print(result.samples)  # [(0, 1, 2, 1, 2), (2, 0, 3, 0, 1), ...
\end{lstlisting}
    \end{figure*}
    
    \begin{figure*}[htbp]
\begin{lstlisting}[
            language=Python,
            caption={The definition of a single CVQNN layer using Piquasso on a single mode. By running the program using the \lstinline{simulator} inside a \lstinline{GradientTape} context, one can extract a quantity from the circuit defined in \lstinline{program} which can be automatically differentiated.},
            label={code:tf_general}
        ]
import numpy as np
import piquasso as pq
import tensorflow as tf

cutoff = 7  # Fock space truncation
d = 1  # Number of modes

target_state_vector = tf.constant(
    [0., 1./np.sqrt(2), 1./np.sqrt(2), 0., 0., 0., 0.], 
    tf.float64,
)

# The neural network weights, initialized with 0
weights = tf.Variable([0., 0., 0., 0.], dtype=tf.float64)

with tf.GradientTape() as tape:
    # Program definition with single CV neural
    # network layer
    with pq.Program() as program:
        pq.Q() | pq.Vacuum()
        pq.Q() | pq.Squeezing(weights[0])
        pq.Q() | pq.Phaseshifter(weights[1])
        pq.Q() | pq.Displacement(weights[2])
        pq.Q() | pq.Kerr(weights[3])
    
    # Enabling automatic differentiation via `TensorflowConnector`
    simulator = pq.PureFockSimulator(
        d=d,
        config=pq.Config(cutoff=cutoff),
        connector=pq.TensorflowConnector()
    )
    
    # Simulating `program` using `simulator`
    state = simulator.execute(program).state
    state_vector = state.state_vector

    # Calculating the cost function `J`
    J = tf.math.sqrt(tf.reduce_sum(tf.abs(target_state_vector - state_vector)**2))

# Automatically differentiating the cost
# function by the CV neural network weights
gradients = tape.gradient(J, weights)
\end{lstlisting}
    \end{figure*}

    \begin{figure*}[htbp]
\begin{lstlisting}[
            language=Python,
            caption={Example usage of \lstinline{SamplingSimulator} using losses with $95 \%$ transmissivity modeled by the \lstinline{Loss} channel.},
            label={code:bosonsamplingexample_with_losses}
        ]
import numpy as np
import piquasso as pq

with pq.Program() as program:
    pq.Q(all) | pq.StateVector([1, 2, 0, 2, 1])
    
    pq.Q(1,2) | pq.Beamsplitter(theta=np.pi / 3)
    pq.Q(2,3) | pq.Beamsplitter(theta=np.pi / 4)
    pq.Q(0,1) | pq.Beamsplitter(theta=np.pi / 5)
    pq.Q(3,4) | pq.Beamsplitter(theta=np.pi / 6)

    for i in range(5):
        pq.Q(i) | pq.Loss(transmissivity=0.95)

    pq.Q(all) | pq.ParticleNumberMeasurement()

simulator = pq.SamplingSimulator(d=5)
result = simulator.execute(program, shots=1000)
print(result.samples)  # [(0, 1, 2, 1, 2), (2, 0, 3, 0, 1), ...
\end{lstlisting}
    \end{figure*}

    \begin{figure*}[htbp]
\begin{lstlisting}[
            language=Python,
            caption={Basic example of using Piquasso \lstinline{PureFockSimulator} with Tensorflow \lstinline{tf.function}. The example uses the \lstinline{pq.cvqnn} module, which implements CVQNN layers. The \lstinline{calculate_cost} function calculates the cost described by Eq.~\eqref{eq:cost} using a randomly generated target state vector. Note, that \lstinline{TensorflowConnector} is supplied with the constructor argument \lstinline{decorate_with=tf.function} in order to decrease the compilation time of the tracing step.},
            label={code:tffunction}
        ]
import scipy
import numpy as np
import piquasso as pq
import tensorflow as tf


@tf.function
def calculate_cost(psi_target, w, cutoff):
    d = pq.cvqnn.get_number_of_modes(w.shape[1])

    simulator = pq.PureFockSimulator(
        d, pq.Config(cutoff=cutoff),
        connector=pq.TensorflowConnector(decorate_with=tf.function),
    )

    with tf.GradientTape() as tape:
        cvqnn_layers = pq.cvqnn.create_layers(w)

        program = pq.Program(instructions=[pq.Vacuum()] + cvqnn_layers.instructions)

        simulator.execute(program)

        psi = simulator.execute(program).state.state_vector

        cost = tf.math.sqrt(tf.math.reduce_sum(tf.math.abs(psi - psi_target)**2))

    return cost, tape.gradient(cost, w)

number_of_layers = 3
d = 4
cutoff = 7

w = tf.Variable(pq.cvqnn.generate_random_cvqnn_weights(number_of_layers, d))

# Size of the state vector in the truncated Fock space
state_vector_size = scipy.special.comb(d + cutoff - 1, cutoff - 1, exact=True)

psi_target = (
    np.random.rand(state_vector_size)
    + 1j * np.random.rand(state_vector_size)
)

# Normalization
psi_target /= np.sum(np.abs(psi_target))

cost, cost_grad = calculate_cost(psi_target, w, cutoff)
\end{lstlisting}
    \end{figure*}

    \begin{figure*}[htbp]
\begin{lstlisting}[
            language=Python,
            caption={Defining a custom simulator class with customized functions for simulation. The \lstinline{_state_class} attribute serves to specify a custom state class that overrides \lstinline{State}, and it is strictly required to specify one. In addition, one can optionally specify a \lstinline{Config} class with the \lstinline{_config_class} class attribute. Some calculations (like hafnian calculation) can be replaced by overriding, e.g., \lstinline{NumpyConnector} and setting the \lstinline{_connector_class} attribute. Finally, one can specify calculation functions for custom instructions to be executed with the \lstinline{_instruction_map} attribute. 
            This code does not serve as a working example, only as a schematic guide.
            },
            label={code:customsimulator}
        ]
class CustomSimulator(pq.Simulator):
    _state_class: Type[State] = CustomState

    _config_class: Type[Config] = CustomConfig

    _connector_class: Type[BaseConnector] = CustomConnector

    _instruction_map: Dict[Type[Instruction], Callable] = {
        CustomPreparation: calculate_custom_preparation,
        CustomGate: calculate_custom_gate,
        CustomMeasurement: calculate_custom_measurement,
    }
\end{lstlisting}
    \end{figure*}

    \begin{figure*}[htbp]
\begin{lstlisting}[
            language={Python},
            caption={Specifying the custom loop hafnian function that is injected into Piquasso. During simulations, the newly created \lstinline{CustomConnector.loop_hafnian} function will be executed instead of the builtin loop hafnian function.},
            label={code:customloophafnian}
        ]
import piquasso as pq

def custom_loop_hafnian_function(matrix, reduce_on):
    """Custom loop hafnian implementation goes here"""
    
# Creating a Connector with the newly created function for computing loop_hafnians
class CustomConnector(pq.NumpyConnector):
    def loop_hafnian(self, matrix, reduce_on):
        return custom_loop_hafnian_function(matrix, reduce_on)

# Creating the custom simulator instance
simulator = pq.GaussianSimulator(d=5, connector=CustomConnector())
\end{lstlisting}
    \end{figure*}

    \begin{figure*}[htbp]
\begin{lstlisting}[
            language={Python},
            caption={A schematic usage example of the PiquassoBoost plugin.},
            label={code:piquassoboost},
        ]
import piquasso as pq
import piquassoboost as pqb
from scipy.stats import unitary_group

d = 20
interferometer_matrix = unitary_group.rvs(d)

# Regular Piquasso program
with pq.Program() as program:
    # Apply squeezing gates to all modes
    for i in range(d):
        pq.Q(i) | pq.Squeezing(r=0.9)

    # Apply a random interferometer
    pq.Q(all) | pq.Interferometer(interferometer_matrix)

    # Measure on all modes
    pq.Q(all) | pq.ThresholdMeasurement()

# PiquassoBoost simulator
boosted_simulator = pqb.BoostedGaussianSimulator(d)
result = boosted_simulator.execute(program)
\end{lstlisting}
    \end{figure*}

\end{document}